\documentclass[a4paper,11pt]{article}
\pdfoutput=1 

\usepackage{jcappub} 
\usepackage{bm,graphicx,dcolumn,epstopdf,epsf, latexsym, mathbbol, amssymb,amsmath,slashed, mathrsfs,mathcomp,simplewick}
\usepackage[center]{subfigure}
\usepackage{multirow}
\usepackage{makecell}
\usepackage{color}

	\newcommand{\bq}{\begin{equation}}
	\newcommand{\eq}{\end{equation}}
	\newcommand{\bqn}{\begin{eqnarray}}
	\newcommand{\eqn}{\end{eqnarray}}
	\newcommand{\nb}{\nonumber}
	\newcommand{\lb}{\label}

	\newcommand{\red}{\textcolor{black}}

\title{Spherical Accretion Flow onto General Parameterized Spherically Symmetric Black Hole Spacetimes}

\author[a,b]{Sen Yang,}
\author[b,c]{Cheng Liu,}
\author[b, c, 1]{Tao Zhu, \note{Corresponding author.}}
\author[a]{Li Zhao,}
\author[b,c]{ Qiang Wu,}
\author[d]{Ke Yang,}
\author[b,c,e]{ Mubasher Jamil}

\affiliation[a]{Institute of theoretical physics, Lanzhou University,\\Lanzhou 730000, China}
\affiliation[b]{Institute for theoretical physics and Cosmology, Zhejiang University of Technology,\\Hangzhou 310032, China}
\affiliation[c]{United center for gravitational wave physics (UCGWP), Zhejiang University of Technology,\\Hangzhou 310032, China}
\affiliation[d]{School of Physical Science and Technology, Southwest University,\\Chongqing 400715, China}
\affiliation[e]{School of Natural Sciences, National University of Sciences and Technology,\\Islamabad, 44000, Pakistan}

\emailAdd{yangs19@lzu.edu.cn}
\emailAdd{liucheng@zjut.edu.cn}
\emailAdd{zhut05@zjut.edu.cn}
\emailAdd{lizhao@lzu.edu.cn}
\emailAdd{wuq@zjut.edu.cn}
\emailAdd{keyang@swu.edu.cn}
\emailAdd{mjamil@zjut.edu.cn}

\abstract{The transonic phenomenon of black hole accretion and the existence of the photon sphere are the characteristics of strong gravitational fields near a black hole horizon. In this work, we study spherical accretion flow onto a general parametrized spherically symmetric black hole spacetimes. For this purpose, we analyze the accretion process of various perfect fluids, such as the isothermal fluid of ultra-stiff, ultra-relativistic, and sub-relativistic types and polytropic fluid, respectively. The influences of extra parameters beyond the Schwarzschild black hole in the general parameterized spherically symmetric black hole on the flow behaviors of the above-mentioned test fluids are studied in detail. In addition, by studying the accretion of ideal photon gas, we further discuss the correspondence between the sonic radius of accreting photon gas and the photon sphere for the general parameterized spherically symmetric black hole. Some possible future extensions of our analysis are also discussed.}

                                           
 \arxivnumber{2006.04715}

\begin{document}

\maketitle
\flushbottom

\section{Introduction}
\renewcommand{\theequation}{1.\arabic{equation}} \setcounter{equation}{0}
\label{sec:intro}
Accretion process around a massive gravitational object is a basic phenomenon in astrophysics and has played essential roles in understanding various astrophysical processes and observations, including the growth of stars, the formation of supermassive black holes, luminosity of quasar, and X-ray emission from compact star binaries, etc (see \cite{frank, yuan, bambi} and references therein). The accretion of matter in a realistic astrophysical process is rather complicated, since it involves a lot of challenging issues of general relativistic magnetohydrodynamics, including turbulence, radiation processes, nuclear burning and more. To understand the general accretion processes, it is powerful to simplify the problem by some assumptions or considering some simple scenarios. 

The simplest accretion scenario consists of the stationary, spherically symmetric solution first discussed by Bondi \cite{bondi1952}, where an infinitely large homogeneous gas cloud steadily accreting on to a central gravitational object was considered. Bondi's treatment is formulated in the framework of Newtonian gravity. Later, in the framework of general relativity (GR), the steady-state spherically symmetric flow of test fluids onto a Schwarzschild black hole was investigated by Michel \cite{michel}. Since then, the spherical accretion has been analyzed in literatures for various spherically symmetric black holes in GR and modified gravities, see \cite{S. K. Chakrabarti, Astrophysical Flows, accretion1, accretion2, accretion3,accretion4,accretion5,accretion6,accretion7,accretion8,accretion9,accretion10,accretion11, accretion12, accretion13, accretion14, accretion15, accretion16, Ahmed:2016ucs, Bahamonde:2015uwa,C. Bambi,Yang:2018cim,Jiao:2016uiv,Yang:2015sfa,Chinese.P.C} and references therein for examples.

One important feature of spherical accretion onto black holes is the phenomenon of transonic accretion and the existence of the sonic point (or critical point). At sonic point, the accretion flow transits from subsonic to supersonic state. Normally, the locations of the sonic points in a given black hole spacetime are not far from its horizon. What ever important and intriguing, the narrow region around sonic point is closely related to ongoing observations in the electromagnetic and gravitational waves spectra. Therefore, the study of the spherical accretion problem can not only help us to understand the accretion process in different black holes, but also importantly, provide an alternative approach for us to explore the nature of the black hole spacetimes in the regime of strong gravity.

On the other hand, the EHT Collaboration recently announced their first image concerning the detection of shadow of a supermassive black hole at the center of a neighboring elliptical M87 galaxy \cite{m87, Akiyama:2019brx, Akiyama:2019eap, Akiyama:2019bqs, Akiyama:2019fyp, Akiyama:2019sww}. With this image, it is observed that the diameter of the center black hole shadow is $(42\pm 3) \; {\rm \mu as}$ which leads to a measurement of the center mass of $M=(6.5\pm 0.7)\times 10^9 M_{\odot}$ \cite{m87}. The outer edge of the shadow image, if one considers a Schwarschild black hole, forms a photon sphere near the black hole horizon, at which the trajectories of photon take a closed circular orbit. Within astrophysical observations, the existence of a photon sphere is related to the electromagnetic observations of black holes through background electromagnetic emission and the frequencies of quasi-normal modes. The later is determined by the parameters of null geodesic motions on and near the photon sphere of a given black hole spacetime. 

Recently, it was shown that there is a correspondence between the sonic points of accreting ideal photon gas and the photon sphere in static spherically symmetric spacetimes \cite{Koga:2016jjq}. This important result is valid not only for spherical accretion of the idea photon gas, but also for rotating accretion in static spherically symmetric spacetimes \cite{Koga:2019teu, Koga:2018ybs}. In an observational viewpoint, as mentioned in \cite{Koga:2018ybs}, this correspondence connects two independent observations, the observation of lights from sources behind a black hole and the observation of emission from accreted radiation fluid onto the black hole, because the size of the shadow of the hole is determined by the radius of the photon sphere and the accreted fluid can signal the sonic point. 

With the above mentioned motivations, \red{it is interesting to explore that spherical accretion flow in different black hole spacetime. The extra parameters beyond the Schwarzschild black hole in these spacetime may affect the flow behaviors of the accretion, thus it provides an important approach to study the   strong gravity behavior of black holes in a lot of alternative theories of gravity. Instead of finding the exact solution and studying the spherical accretion case by case in each given theory, a reasonable strategy is to consider a model- independent framework that parametrizes the most generic black-hole geometry through a finite number of adjustable quantities. For this purpose,} in this paper, we consider the spherical accretion flow in a general parameterized spherically symmetric black hole spacetimes \cite{rz}. This parameterized description allows one to consider the accretion phenomena not in each specific theories of gravity, but in a unified way by exploring the influence of different black hole parameters on the spherical accretion process \cite{rz}.  Specifically, we focus our attention on perfect fluid accretion on to the general parameterized spherically symmetric black hole spacetimes and study in details the transonic phenomenon for different types of fluid, including the isothermal fluids and polytropic fluid. By studying the accretion of idea photon gas, we further reveal the correspondence between the sonic points of accreting photon gas and the photon sphere for the general parameterized spherically symmetric black hole.

The plan of our paper is as follows. In Sec. II, we present a very brief introduction of the general parameterized spherically symmetric black hole. Then in Sec. III we derive the basic equations for later discussions of the spherical accretion of various fluids and gives several useful quantities. Sec. IV is devoted to performing a dynamical system analysis of the accretion process and finding the critical points of the system. With these results, in Sec. V we apply these obtained formalism or results to several known fluids and study in details the transonic phenomenon for accretion of these fluids in the general parameterized spherically symmetric black hole. In Sec. VI, by studying the spherical accretion of ideal photon gas and photon sphere of the general parameterized spherically symmetric black hole, we establish the correspondence between the sonic points of ideal photon gas and its photon sphere. The conclusion of this paper is presented in Sec. VII.

\section{Parameterized spherically symmetric black hole spacetime }
\renewcommand{\theequation}{2.\arabic{equation}} \setcounter{equation}{0}

In this section, we present a brief introduction of the \red{L. Rezzolla and A. Zhidenko's (RZ)} parameterization \cite{rz} for a generic spherically symmetric black hole spacetime. To start, let us first consider the line element of any spherically symmetric stationary configuration in a spherical polar coordinate system $(t, r, \theta, \phi)$, which can be written as
\bqn\lb{matric}
ds^2 = - N^2(r) dt^2 + \frac{B^2(r)}{N^2(r)}dr^2 + r^2 (d\theta^2 + \sin^2\theta d\phi^2),
\eqn
where $N(r)$ and $B(r)$ are two functions of the radial coordinate $r$ only. In the RZ parameterization, $N(r)$ is expressed in the form
\bqn\lb{N}
N^2(x) = x A(x),
\eqn
where $A(x) >0$ for $0<x<1$ with $x = 1- r_0/r$. It is obvious that $x=0$ represents the location of the event horizon of the black hole and $x=1$ is the spatial infinity. Then the function $A(x)$ and $B(x)$ can be further parameterized in terms of the parameters $\epsilon$, $a_i$, and $b_i$ as
\bqn\lb{A(x)}
A(x) &=& 1 - \epsilon(1-x) +(a_0-\epsilon)(1-x)^2+ \tilde{A}(x) (1-x)^3, \\ 
B(x) &=& 1 + b_0 (1-x) + \tilde{B}(x) (1-x)^2,\lb{B(x)}
\eqn
where the functions $\tilde{A}$ and $\tilde{B}$ are introduced to describe the metric near the horizon (i.e., $x \simeq 0$) and at the spatial infinity (i.e., $x=1$). \red{The coefficients $a_0$ and $b_0$
can be seen as combinations of the PPN parameters.} The functions $\tilde A$ and $\tilde B$ can be expanded by the continuous  Padé approximation as
\bqn
\tilde A(x) = \frac{a_1}{1+ \frac{a_2 x}{1+ \frac{a_3 x}{1+ \cdots }}},\;\;\; \tilde B(x) = \frac{b_1}{1+ \frac{b_2 x}{1+ \frac{b_3 x}{1+ \cdots }}},
\eqn
where $a_1, a_2, \cdots, a_n$ and $b_1, b_2, \cdots, b_n$ are dimensionless constants that can be determined by matching the above parameterization to a specific metric. In addition, the parameter $\epsilon$ in the RZ parameterization measures the deviations of the position of the event horizon in the general metric from the corresponding location in a Schwarzschild spacetime, i.e.,
\bqn
\epsilon = \frac{2 M - r_0}{r_0}.
\eqn

The RZ parameterization can be matched to a lot of black hole solutions which differ from GR. These solutions, just mentioned a few, include Reissner-Nordstr\"{o}m (RN) black hole in GR and black holes in Brans-Dicke gravity (BD), $f(R)$ gravity, Einstein-Maxwell-Axion-Dilaton theory (EMAD), Einstein-\AE{}ther theory, etc \cite{rz, rz2}. Recently, RZ parameterization has also been extended to the rotating case \cite{krz}.

\section{Basic equations for spherical accretion flow}
\renewcommand{\theequation}{3.\arabic{equation}} \setcounter{equation}{0}

In this section, we start to consider the steady state spherical accretion flow of matter near the RZ parameterized black hole. For this purpose, the accreting matter is approximated by a relativistic perfect fluid with neglecting effects related to viscosity or heat transport. In this way, the energy momentum tensor of the fluid can be described by 
\begin{equation}\lb{Tmn}
T^{\mu \nu}=(\rho+p)u^\mu u^\nu + p g^{\mu\nu},
\end{equation}
where $\rho$ and $p$ are the proper energy density and the pressure of the perfect fluid. The four velocity $u^\mu$ obeys the normalization condition $u_\mu u^\mu = -1$. We assume that the fluid is radially flowing into the black holes, therefore we have $u^\theta = 0 = u^\phi$. For the same reason, the physical quantities ($\rho$, $p$) and others introduced later are functions of the radial coordinate $r$ only. For the sake of simplicity, we set the radial velocity as $u^r = u <0$ for the accreting case. Then using the normalization condition, it is easy to infer that 
\bqn
(u^t)^2 =\frac{N^2(r)+B^2(r) u^2}{N^4(r)}.  
\eqn

There are two basic conservation laws which govern the evolution of the fluid in the black hole spacetime. One is the conservation law of the particle number, and another one is the conservation of the energy momentum. The assumption of the conservation of the particle number implies there is no particle creation and annihilation during the accreting process. Defining the proper particle number density $n$ and number current $J^\mu = n u^\mu$ in the local inertial rest frame of the fluid, the conservation of the particle number gives
\bqn\lb{particle}
\nabla_\mu J^\mu = \nabla_{\mu} (n u^\mu)=0,
\eqn
where $\nabla_\mu$ denotes the covariant derivative with respect to the coordinate.  For the RZ parameterization of a generic spherically symmetric black hole spacetime, Eq.~(\ref{particle}) can be rewritten as 
\bqn\lb{ncurr}
\frac{1}{r^2 B} \frac{d}{d r}(r^2 B n u) = 0.
\eqn
Integrating this equation gives, 
\bqn\lb{Eq1}
r^2 B n u = C_1,
\eqn
where $C_1$ is the integration constant. 

The conservation law of the energy momentum is expressed as
\bqn\lb{EMC}
\nabla_\mu T^{\mu \nu} =0.
\eqn
It is also convenient to introduce the first law of the thermodynamics of the perfect fluid, which is described by \cite{book1}
\bqn\lb{theom}
dp = n (dh - T ds),~~~d\rho = h dn +nT ds,
\eqn
where $T$ is the temperature, $s$ is the specific entropy and $h$ is the specific enthalpy defined as
\bqn
h \equiv \frac{\rho + p}{n}. \lb{hh}
\eqn
Then projecting conservation law of the energy-momentum (\ref{EMC}) along $u^\mu$, one obtains
\bqn
u_{\nu}\nabla_{\mu} T^{\mu\nu} &=& u_{\nu} \nabla_{\mu} \Big[ nh u^{\mu} u^{\nu} + p g^{\mu\nu}\Big] \nb\\
&=& - n u^{\mu} \nabla_{\mu} h + u^{\mu} \nabla_{\mu} p.
\eqn
In above we have used the conservation of the particle number, i.e., $\nabla_{\mu} (n u^{\mu}) =0$ and $u^{\mu} \nabla_{\nu} u_{\mu} =u_{\mu} \nabla_{\nu} u^{\mu} = \frac{1}{2} \nabla_{\nu} (u^{\mu}u_{\mu})=0$. Noticing that the first thermodynamical law (\ref{theom}) can be rewritten as $\nabla_{\mu  }p = n \nabla_{\mu} h - n T \nabla_{\mu}s$, from the above projection one arrives 
\bqn
- n T u^{\mu} \nabla_{\mu} s =0,
\eqn
implying that there is no heat transfer between the different fluid elements and  the specific entropy is conserved along the evolution lines of the fluid.  In the parameterized spherically symmetric black hole, the conservation of the specific entropy reduces to $\partial_r s=0$, i.e., \red{$s={\rm constant}$.} For this reason, the fluid is isentropic and Eq.~(\ref{theom}) reduces to
\bqn
dp = n dh,~~~~d\rho = h dn.
\eqn

With the above thermodynamical properties of the perfect fluid, the conservation law of the energy-momentum (\ref{EMC}) can be written as
\bqn
\nabla_\mu T^{\mu}_{\nu} &=& \nabla_{\mu} (h n u^\mu u_\nu) +\nabla_{\mu} (\delta^\mu_\nu p)\nb\\
&=& n u^{\mu} \nabla_{\mu} (h u_\nu) +  n \nabla_{\nu }h \nb\\
&=& n u^{\mu} \partial_{\mu} (h u_\nu) - n u^{\mu} \Gamma_{\mu \nu}^\lambda h u_{\lambda}+  n \nabla_{\nu }h\nb\\
&=& 0.
\eqn
Then the time component $\nu= t$ of above equation yields
\bqn
\partial _r(hu_t) = 0.
\eqn
Integrating it for the parameterized spherically symmetric black hole we considered in this paper, one arrives at
\bqn\lb{Eq2}
h \sqrt{N^2 +B^2u^2} = C_2,
\eqn
where $C_2$ is the integration constant. This equation, together with Eq.~(\ref{Eq1}), constitutes the two basic equations describing a radial, steady-state perfect fluid flow in the parameterized spherically symmetric black hole. 

To proceed further, let us introduce several useful quantities to describe the accretion flow, which will be used in the later analysis. The first quantity is the sound speed of the perfect fluid, which is defined by
\bqn
c_s^2 \equiv \frac{d p}{d \rho} = \frac{n}{h}\frac{d h}{d n} = \frac{d \ln h }{d \ln n}. \lb{cs}
\eqn
On the other hand, by considering accretion flow is radial, i.e., $ d \theta = d \phi = 0$, the black hole metric can be decomposed as\cite{Azreg-Ainou:2018wjx}
\bqn
ds^2 = - (N dt)^2 + \left(\frac{B}{N}dr \right)^2,
\eqn
from which one can define an ordinary three-dimensional velocity $v$ measured by static observers as
\bqn
v \equiv \frac{B}{N^2} \frac{dr}{dt}.
\eqn
Considering $u^{r}= u = dr/d\tau$ and $u^t = dt /d\tau$ with $\tau$ being the proper time of the fluid, one finds
\bqn
v^2 =  \frac{B^2}{N^4} \left(\frac{u}{u^t}\right)^2 = \frac{B^2 u^2}{N^2+B^2 u^2}.
\eqn
Then one can express $u^2$ and $u_t^2$ in terms of $v^2$ as
\bqn\lb{utov}
u^2 = \frac{N^2 v^2}{B^2(1-v^2)},
\eqn
\bqn
u^2_t = \frac{N^2}{1-v^2}.
\eqn
These quantities will be used in the following dynamical system analysis for the radial, steady-state perfect fluid flow in the parametrized spherically symmetric black hole. 

\section{Sonic Points and Dynamical system analysis}
\renewcommand{\theequation}{4.\arabic{equation}} \setcounter{equation}{0}

The two basic equations (\ref{Eq1}) and (\ref{Eq2}) consist a dynamical system for the radial accretion process. In this section, we use these equations to study the accretion process in the parametrized spherically symmetric black hole. 

\subsection{Sonic points}

In the trajectories of the accretion flow into the black hole, there exists a specific point called sonic point, at which the four-velocity of the moving fluid becomes equal to the local speed of sound and the accretion flow has the maximum accretion rate. To determine the sonic point, let us first take the derivative of the two basic equations  (\ref{Eq1}) and (\ref{Eq2}) with respect to $r$, which leads to
\bqn
&&\left(v^2 - c_s^2\right) \frac{d \ln v}{dr} = \frac{1-v^2}{B N r} \left[ c_s^2 N B \left(2  +r \frac{d\ln B}{dr}\right)- B (1-c_s^2) r \frac{dN}{dr}\right].
\eqn
At the sonic point $r_*$ ($c_s^2(r_*)=v^2(r_*)$), one has
\bqn
c_{s*}^2 N_* B_* \left(2  +r_* \left.\frac{d\ln B}{dr}\right|_*\right)- B_* (1-c_{s*}^2) r_* \left.\frac{dN}{dr}\right|_*=0,
\eqn
where the $*$ denotes the values evaluated at the sonic point. This equation allows us to determine the sonic point once the speed of sound $c_s^2 \equiv dp/d\rho$ is known. The above equation can be rewritten as
\bqn
u_*^2 = \frac{N_* r_*\left.\frac{dN}{dr}\right|_*}{B_*^2 \left(2  +r_* \left.\frac{d\ln B}{dr}\right|_*\right)}.
\eqn
Therefore, once $r_*$ is determined, one can use this expression to find the value of $u$ at the sonic point. The existence of the sonic point in black hole spacetime physically exhibits a very interesting accreting phenomenon, that it highlights the transonic solutions which are supersonic near and subsonic far from the black hole. In the following sections, we are going to find sonic points by using the equations obtained in this subsection and discuss the transonic phenomenon in details for different fluids.

\subsection{Dynamical system and critical points}

From the two basic equations  (\ref{Eq1}) and (\ref{Eq2}), we observe that there are two integration constants $C_1$ and $C_2$. For this kind of system,  we may treat the square of the left-hand side of Eq.~(\ref{Eq2}) as a Hamiltonian $\mathcal{H}$ of this system,
\bqn
\mathcal{H} =h^2 (N^2+B^2 u^2),
\eqn
so $C_2$ of every orbit in the phase space of this system is kept fixed. 
Inserting Eq.~(\ref{utov}) into the Hamiltonian $\mathcal{H}$ one finds
\bqn\lb{H(r,v)}
\mathcal{H}(r,v) = \frac{h^2(r,v)N^2}{1-v^2}.
\eqn
Then the dynamical system associated with this Hamiltonian reads
\bqn
\dot{r} = \mathcal{H}_{,v} ,~~~~\dot{v} = -\mathcal{H}_{,r},
\eqn
where the  dot denotes the derivative with respect to $\bar t$ with being the time variable of the Hamiltonian dynamical system. Then inserting the Hamiltonian one finds
\bqn
\dot{r} &\equiv& f(r,v) = \frac{2 h^2 N^2}{v(1-v^2)^2} (v^2 - c_s^2), \lb{DY1} \\ 
\dot{v} &\equiv& g(r,v)= -\frac{h^2 }{r (1-v^2)} \left[r N^2_{,r} (1 - c_s^2) - 4 N^2 c_s^2\right]. \lb{DY2}
\eqn
These equations constitute an autonomous, Hamiltonian two-dimensional dynamical system. Its orbits are composed of the solutions of the two basic equations (\ref{Eq1}) and (\ref{Eq2}). \red{In the construction of  the above dynamical system, we consider the two quantities $(r, v)$ as the two dynamical variables of the system. It is worth to mention that there are actually different ways to fix the dynamical variables, for examples, one may choose the dynamical variables to be $(r, h)$, $(r, p)$, or $(r, u)$ \cite{Ahmed:2015tyi}. }

At the critical points, the right-hand sides of Eqs.~(\ref{DY1}) and (\ref{DY2}) vanish, and the following equations provide a set of critical points that are solutions to $\dot{r} = 0$ and $\dot{v} = 0$,
\bqn
v_*^2 &=& c_s^2, \\
c_s^2 &=& \frac{r_* N^2_{*,r_*}}{r_* N^2_{*,r_*} + 4N^2_{*,r_*}} .\lb{c_s^2}
\eqn
It is clear to see that sonic points are the critical points of this dynamical system. Hereafter we use $(r_*, v_*)$ to denote the critical points of the dynamical system. For a dynamical system, the critical points can be divided into several different types. In order to see what types of critical point could arise from the black hole accreting process, let us perform the flowing linearization of the dynamical system by Taylor expanding of Eqs.~(\ref{DY1}) and (\ref{DY2}) around the critical points, i.e.,   
\bqn\lb{det}
\left( \begin{array}{cccc}
	{\delta \dot r} \\
	{\delta \dot v}
\end{array}\right)
=
X
\left( \begin{array}{cccc}
	\delta r \\
	\delta v
\end{array}\right),
\eqn
where $\delta r$, $\delta v$ denote the small perturbations of $r$, $v$ about the critical points and $X$ is the Jacobian matrix of the dynamical system at the critical point $(r_*, v_*)$, which is defined as
\bqn
X=
\left( \begin{array}{cccc}
	\frac{\partial{f}}{\partial{r}} & \frac{\partial{f}}{\partial{v}}   \\
	\frac{\partial{g}}{\partial{r}} & \frac{\partial{g}}{\partial{v}}
\end{array}\right)\Bigg|_{(r_*, v_*)}.
\eqn
Depending on the determinant $\Delta = {\rm det} (X)$ of $X$ and its trace $ \chi = {\rm Tr}(X)$, the types of the critical points $(r_*, v_*)$ of the dynamcial system can be summarized as follows,
\begin{itemize}
	\item Saddle points if $\Delta <0$.
	\item Attracting nodes if $\Delta >0$, $\chi < 0$, and $\chi^2-4\Delta >0$. 
	\item Attracting spirals if $\Delta >0$, $\chi < 0$, and $\chi^2-4\Delta <0$.  
	\item Repelling nodes if $\Delta >0$, $\chi > 0$, and $\chi^2-4\Delta >0$.
	\item Repelling spirals if $\Delta >0$, $\chi > 0$, and $\chi^2-4\Delta <0$.
	\item Degenerate nodes if $\Delta >0$, and $\chi^2-4\Delta =0$.
	\item Centers if $\Delta >0$, $\chi = 0$. 
	\item Line or plane critical points if $\Delta =0$.
\end{itemize}

When the critical points and their types are determined, the constant $C_1$ in Eq.~(\ref{Eq1}) can be rewritten in terms of quantities evaluated at the critical point $(r_*, v_*)$ as
\bqn
C^2 _1 = \frac{r^4 _* n^2 _* v^2 _* N^2_*}{1 -v^2 _*} = \frac{r^5 _* n^2 _* N^2_{*,r_*}}{4}.
\eqn
This equation is satisfied \red{not only at the critical point, but also at any points in the same streamline in the phase portrait,} so one can easily get
\bqn\lb{(n/n_*)^2}
\left(\frac{n}{n_*}\right)^2 = \frac{r^5_* N^2_{*,r_*}}{4}\frac{1- v^2}{r^4 N^2 v^2}.
\eqn 
If there is no solution to Eqs.~(\ref{DY1}) and (\ref{DY2}) at the critical point, one can introduce any reference point $(r_0,v_0)$ from the phase portrait, then arrive at\cite{Ahmed:2016cuy}
\bqn\lb{n^2}
\left(\frac{n}{n_0}\right)^2 = \frac{r^4_0 N^2_0 v^2_0}{1- v^2_0}\frac{1- v^2}{r^4 N^2 v^2}.
\eqn
The above expressions will be used later to analyze the spherical accretion for some test fluids.

\section{Applications to test fluids}
\renewcommand{\theequation}{5.\arabic{equation}} \setcounter{equation}{0}

In this section, we consider the accretion process of several test fluids by using the equations derived in the above sections for the parameterized spherically symmetric black hole. Specifically we consider the isothermal and polytropic fluids respectively in the following subsections.

\subsection{Isothermal test fluid}

In this subsection, we will consider the accretion processes for isothermal fluids, whose temperature is at a constant. This system can be viewed as adiabatic because of the fast speed of fluid. For such systems we define its equation of state (EoS) $w$ as
\begin{equation}
w\equiv p/\rho.
\end{equation}
where $\rho$, $p$ represent the energy density and pressure of the fluid respectively. It is worth noting that $0< w \leq 1$ for isothermal fluid\cite{Mach:2013fsa}. In addition, the adiabatic sound speed is specified by $c_s^2 \equiv \frac{d p}{d \rho} =w $.

According to $h=(\rho+p)/n = (1+w)\rho/n$ and $c_s^2=d\ln h/ d\ln n =w$, we have
\bqn
\rho = \rho_0 \left(\frac{n}{n_0}\right)^{1+w},
\eqn
and
\bqn
h =  \frac{(w+1) \rho_0}{n_0} \left(\frac{n}{n_0} \right)^w,
\eqn
where $n_0$ and $\rho_0$ denote the values of $n$ and $\rho$ evaluated at some reference point. Using Eq.~(\ref{(n/n_*)^2}), we arrive at
\bqn
h^2 = K\left( \frac{1- v^2}{r^4 N^2 v^2}\right)^w.
\eqn
where $K$ is a constant. Through the transformation $\bar{t} \rightarrow K\bar{t}$ and $\mathcal{H} \rightarrow \mathcal{H}/K$, the constant $K$ is absorbed in a redefinition of the time ${\bar{t}}$. Then, the new Hamiltonian becomes 
\bqn\lb{H(N)1}
\mathcal{H}(r,v) = \frac{N^{2(1-w)}}{(1-v^2)^{1-w} v^{2w} r^{4w}}.
\eqn

Considering the first-order in the RZ parameterization \red{(taking only the first three items of Eq. (\ref{A(x)}))}, one can approximately get
\bqn\lb{N2}
N^2 \simeq \left(1-\frac{2M}{r(1+\epsilon)}\right)\left[1+\frac{4 M^2 (a_0- \epsilon)}{r^2 (1+\epsilon)^2} - \frac{2 M \epsilon }{r(1+\epsilon)}\right].
\eqn
Then Eq. (\ref{H(N)1}) can be approximately rewritten as
\bqn\lb{H(w)}
\mathcal{H}
\simeq \frac{\left(1-\frac{2M}{r(1+\epsilon)}\right)^{1-w} \left[1+\frac{4 M^2 (a_0- \epsilon)}{r^2 (1+\epsilon)^2} - \frac{2 M \epsilon }{r(1+\epsilon)}\right]^{1-w}}{(1-v^2)^{1-w} v^{2w} r^{4w}}.
\eqn
At the sonic point, with $c_s^2 = w $, Eq. (\ref{DY2}) reduces to 
\bqn\lb{w(N)}
w = \left.  \frac{r N^2_{,r}}{r N^2_{,r} + 4 N^2} \right|_{r=r_*}.
\eqn   
With Eq.(\ref{N2}), Eq.~(\ref{w(N)}) can be approximately rewritten as
\bqn\lb{w}
w=
\frac{M [-12 M^2 \epsilon + r_*^2 (1 + \epsilon)^3 + 
	4 a_0 M (3 M - r_* (1 + \epsilon))]}{LT_1},
\eqn
where
\bqn
LT_1 = &&4 M^3 \epsilon - 3 M r_*^2 (1 + \epsilon)^3 + 
2 r_*^3 (1 + \epsilon)^3 + 4 a_0 M^2 (-M + r_* + r
_*\epsilon).
\eqn

\subsubsection{Solution for ultra-stiff fluid ($w=1$)}

\begin{figure}
	\centering
	\includegraphics[width=3.4in]{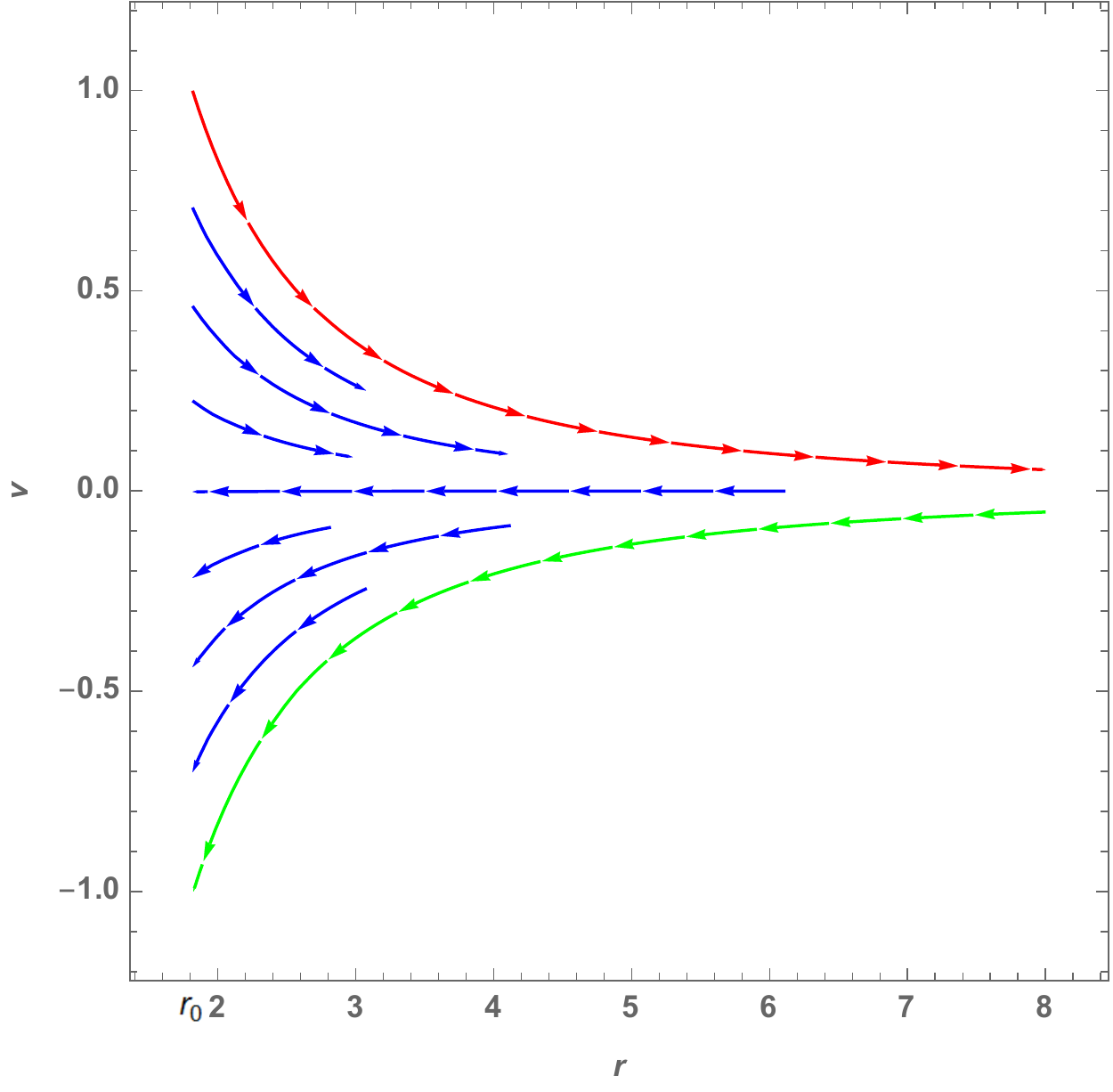}
	\caption{Phase space portraits of the dynamical system (\ref{DY1}, \ref{DY2}) for the ultra-stiff fluid ($w=1$) with black hole  $M=1$, $\epsilon=0.1$, and $a_0=0.0001$.}
	\label{ultra}
\end{figure}

\red{Let us first consider the ultra-stiff fluid, whose energy density is equal to its pressure.} In this case, the equation of state $w=p/\rho=1$. The Hamiltonian (\ref{H(w)}) for the ultra-stiff fluid becomes
\bqn\lb{H(w=1)}
\mathcal{H} = \frac{1}{v^2 r^4}.
\eqn
For physical flows, one has $|v| <1$. Therefore the Hamiltonian (\ref{H(w=1)}) for the ultra-stiff fluid has minimal value $\mathcal{H}_{\rm min} = r_0^{-4}$. With Eq.~(\ref{H(w=1)}), the two-dimensional dynamical system (\ref{DY1}, \ref{DY2}) is
\bqn 
\dot{r} &=&  - \frac{2}{r^4 v^3},\lb{r(w=1)} \\
\dot{v} &=&  \frac{4}{r^5 v^2} .\lb{v(w=1)}
\eqn
It is obvious to see that there is no critical point for this dynamical system. The phase space portraits of this dynamical system for the ultra-stiff fluid with $M=1$, $a_0=0.001$, and $\epsilon=0.1$ of the general parameterized black hole is depicted in Fig.~\ref{ultra}, in which the physical flows of the ultra-stiff fluid in the general parameterized black hole are represented by several curves with arrows. It is shown that the curves with $v<0$ have arrows directed toward into the black hole, which represent the accreting flow of the ultra-stiff fluid, while the curves with $v>0$ have arrows directed toward outside, which represent the outflow fluids. The green and red curves are the flows with minimal Hamiltonian $\mathcal{H}_{\rm min}$ for the accretion outflow respectively. All the flows in Fig.~\ref{ultra} between the red and green curves are physical and have Hamiltonian $\mathcal{H} > \mathcal{H}_{\rm min}$.

The values of the minimal Hamiltonian $\mathcal{H}_{\rm min}$ depends on the parameters in the
Rezzolla-Zhidenko parametrization. In Table.~\ref{tab2}, the values of $r_0$ and $\mathcal{H}$ for different values of parameter $\epsilon$ for the ultra-stiff fluid are presented. Since the horizon radius decreases with respect to $\epsilon$, it is shown clearly that the minimal Hamiltonian $\mathcal{H}_{\rm min} = 1/r_0^4$ increases with the increasing of the parameter $\epsilon$.

\begin{table}[tbp]
	\centering
	\begin{tabular}{|c|c|c|c|c|c|c|}
		\hline
		\textrm{$\epsilon$} & $0$ &$0.1$ & $0.2$ & $0.3$ & $0.4$ & $0.5$ \\
		\hline
		$r_0$ &$2$ & $1.81818$ & $1.66667$ & $1.53846$ &$1.42857$ & $1.33333$ \\
		\hline
		$\mathcal{H}_{min}$& $0.0625$ &$0.0915063$ & $0.1296$ & $0.178506$ & $0.2401$ & $0.316406$ \\
		\hline
	\end{tabular}
	\caption{
		\label{tab2}
		Values of $r_0$ and $\mathcal{H}_{min}$ with different values of the black hole parameter $\epsilon$ for the ultra-stiff fluid $w=1$. In the calculation, we set $M=1$ and $a_0=10^{-4}$.}
\end{table}

\subsubsection{Solution for ultra-relativistic fluid ($w=1/2$)}

Let us turn to the ultra-relativistic fluid, for which we consider the equation of state $w=1/2$, i.e., $p=\rho/2$. \red{In this case, the fluids' isotropic pressure is less than their energy density.} With $w=1/2$, the Hamiltonian (\ref{H(w)}) becomes
\bqn\lb{H(w=1/2)}
\mathcal{H} = \frac{\left(1-\frac{2M}{r(1+\epsilon)}\right)^{1/2} \left[1+\frac{4 M^2 (a_0- \epsilon)}{r^2 (1+\epsilon)^2} - \frac{2 M \epsilon }{r(1+\epsilon)}\right]^{1/2}}{ r^{2}|v|(1-v^2)^{1/2} }, 
\eqn
and then the two-dimensional dynamical system (\ref{DY1}, \ref{DY2}) is
\bqn\lb{r(w=1/2)}
\dot{r} = &&\frac{\left(1-\frac{2M}{r(1+\epsilon)}\right)^{1/2} \left[1+\frac{4 M^2 (a_0- \epsilon)}{r^2 (1+\epsilon)^2} - \frac{2 M \epsilon }{r(1+\epsilon)}\right]^{1/2}}{ r^{2}(1-v^2)^{3/2} } \nb\\
&&-\frac{\left(1-\frac{2M}{r(1+\epsilon)}\right)^{1/2} \left[1+\frac{4 M^2 (a_0- \epsilon)}{r^2 (1+\epsilon)^2} - \frac{2 M \epsilon }{r(1+\epsilon)}\right]^{1/2}}{ r^{2}v^2(1-v^2)^{1/2} },
\eqn
\bqn\lb{v(w=1/2)}
\dot{v} = &&\frac{2\left(1-\frac{2M}{r(1+\epsilon)}\right)^{1/2} \left[1+\frac{4 M^2 (a_0- \epsilon)}{r^2 (1+\epsilon)^2} - \frac{2 M \epsilon }{r(1+\epsilon)}\right]^{1/2}}{ r^{3}|v|(1-v^2)^{1/2} } \nb\\
&& - \frac{\left(1-\frac{2M}{r(1+\epsilon)}\right)^{} \left[-\frac{8 M^2 (a_0- \epsilon)}{r^3 (1+\epsilon)^2} - \frac{2 M \epsilon }{r^2 (1+\epsilon)}\right]^{}}{ LT_2}\nb\\
&&-\frac{2 M (1 + \frac{4 M^2 (a_0 - \epsilon)}{r^2 (1 + \epsilon)^2} - \frac{2 M\epsilon}{r (1 + \epsilon)})}{r^2 (1 + \epsilon) LT_2},
\eqn
where
\bqn
LT_2=&& 2r^{2}|v|(1-v^2)^{1/2} \left(1-\frac{2M}{r(1+\epsilon)}\right)^{1/2} \left[1+\frac{4 M^2 (a_0- \epsilon)}{r^2 (1+\epsilon)^2} - \frac{2 M \epsilon }{r(1+\epsilon)}\right]^{1/2}.
\eqn
For some given value of $\mathcal{H}$, one can obtain $v^2$ from Eq.~(\ref{H(w=1/2)}),
\bqn
v^2 =\frac{1 \pm \sqrt{1+\frac{4F(r)}{r^{4} \mathcal{H}_0^2}}}{2},
\eqn
where 
\bqn
F(r) = &&-1 + \frac{2 M}{r} + \frac{8 M^3}{r^3 (1 + \epsilon)^3} + \frac{8 a_0 M^3}{r^3 (1 + \epsilon)^3} \nb\\&&- \frac{8 M^3}{r^3 (1 + \epsilon)^2} - \frac{4 a_0 M^2}{r^2 (1 + \epsilon)^2}.
\eqn

\begin{figure}
	\centering
	\includegraphics[width=3.4in]{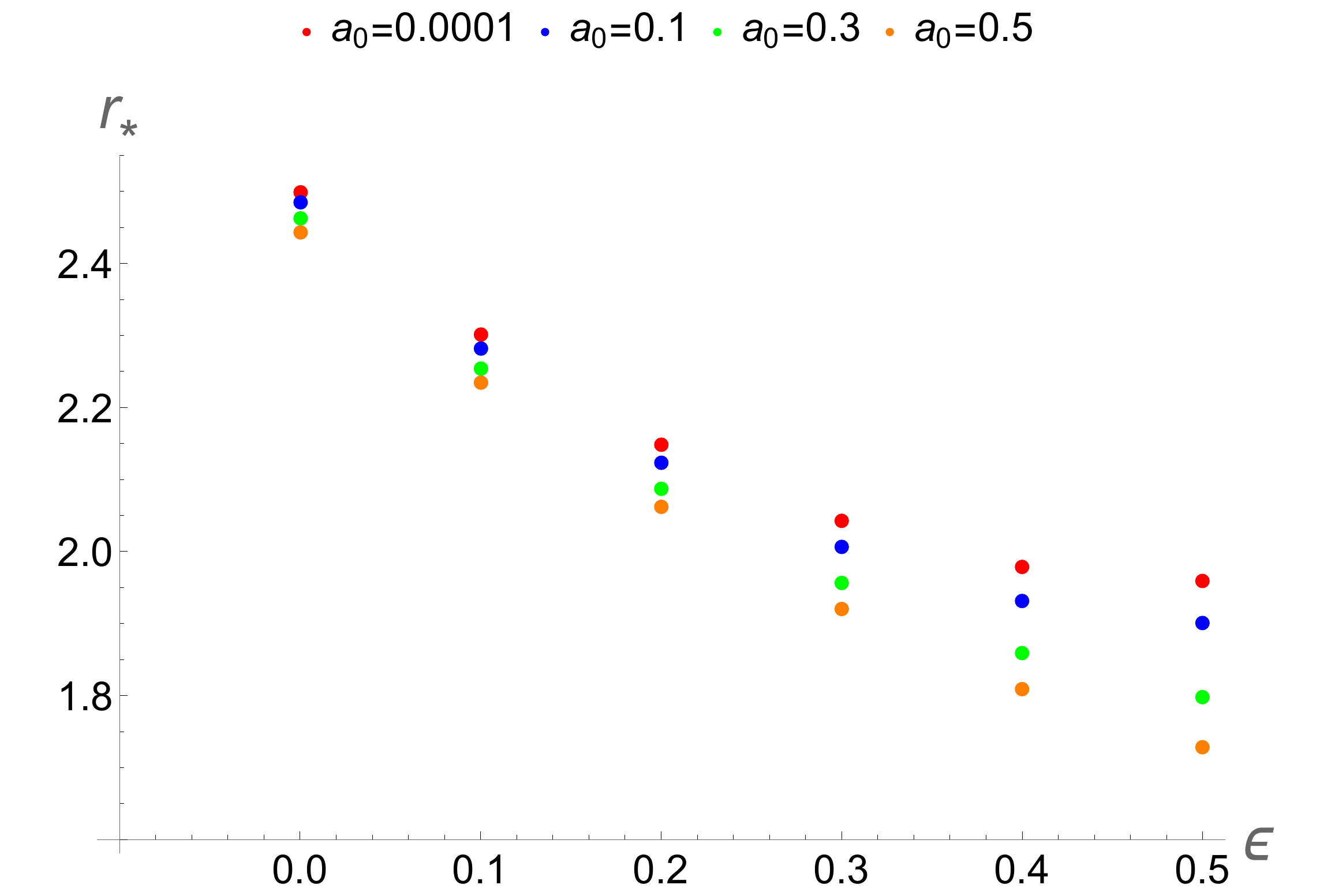}
	\caption{Relation of $r_*$ and $\epsilon$ for different $a_0$ in the spherical accretion process for ultra-relativistic fluid($w=1/2$).}
	\label{rc(w=1/2)}
\end{figure}

Also, one can get critical points of accretion process for ultra-relativistic fluid by solving the two-dimensional dynamical system, when both right hand sides of Eq.~(\ref{r(w=1/2)}) and Eq.~(\ref{v(w=1/2)}) vanish. With black hole parameters being set to $M=1$, $\epsilon = 0.1$, and $a_0 = 0.0001$, one obtains the physical critical points $(r_*, \pm v_*)$, i.e. $(2.30139,-0.707107)$ and $(2.30139,0.707107)$ for outflow and accreting flow respectively. Inserting these values of critical points $(r_*, \pm v_*)$ into Eq.~(\ref{H(w=1/2)}), one finds the critical Hamiltonian $\mathcal{H}_* = 0.160335$. The values of $r_*$, $\pm v_*$, $\mathcal{H}_*$ at the sonic point with different values of the black hole parameters is summarized in Table.~\ref{tab3} for $w=1/2$, $M=1$, and $a_0 = 0.0001$. One can see that, as the value of the black hole parameter $\epsilon$ grows: (1) the value of $r_*$ at the sonic point becomes smaller, while the distance from horizon $r_0$ to critical point becomes larger; (2) the values of velocity $\pm v_*$ at the sonic points are two constants since they equal to the sound speed of the fluid; (3) the value of the Hamiltonian for fluids at the critical points increase. We also display the behaviors of the critical radius $r_*$ with respect to the black hole parameter $\epsilon$ for different values of parameter $a_0$ in Fig. \ref{rc(w=1/2)}, which shows that the critical radius $r_*$ decreases as the increasing of the black hole parameters $\epsilon$ and $a_0$. 

\begin{figure}
	\centering
	\includegraphics[width=3.4in]{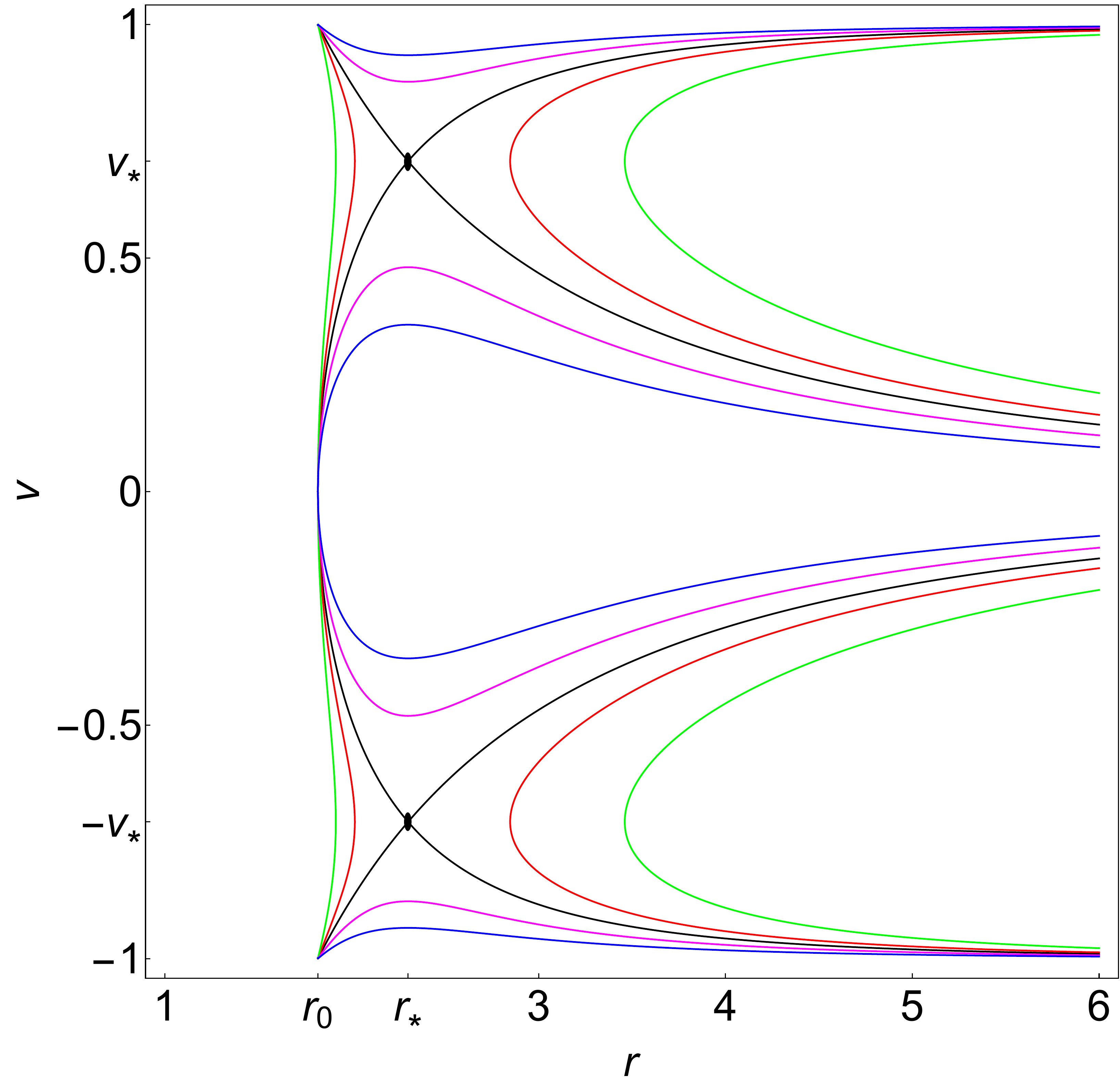}
	\caption{Phase space portraits of the dynamical system (\ref{DY1}, \ref{DY2}) for the ultra-relativistic fluid ($w=1/2$) with black hole parameter $M=1$, $\epsilon=0.1$, and $a_0=0.0001$. The critical (sonic) points $(r_*, \pm v_*)$ of this dynamical system are presented by the black spots in the figure. The five curves with color in black, red, green, magenta, and blue correspond to values of Hamiltonian $\mathcal{H}=\mathcal{H}_*,\; \mathcal{H_*}-0.02, \mathcal{H}_*- 0.05, \mathcal{H}+0.03, \mathcal{H}_* + 0.08$, respectively.}
	\label{ultra-relativistic}
\end{figure}

The phase space portraits of this dynamical system for the ultra-relativistic fluid with $M=1$, $a_0=0.0001$, and $\epsilon=0.1$ of the general parameterized black hole is depicted in Fig.~\ref{ultra-relativistic}, in which the physical flows of the ultra-relativistic fluid in the general parameterized black hole are represents by several curves. One can see that both the critical points in Fig. \ref{ultra-relativistic}, $(r_*, v_*)$ and $(r_*, -v_*)$, are saddle points of the dynamical system. The five curves in Fig. \ref{ultra-relativistic} correspond to different values of the Hamiltonian $\mathcal{H}_0= \{\mathcal{H}_*-0.05,~ \mathcal{H}_*- 0.02,~ \mathcal{H}_*,~\mathcal{H}_*+ 0.03,~\mathcal{H}_*+ 0.08\}$ respectively. This plot shows several different types of fluid motions. The magenta (with $\mathcal{H}=\mathcal{H}_* +0.08$) and blue (with $\mathcal{H}=\mathcal{H}_* +0.03$) curves correspond to purely supersonic accretion ($v <- v_*$ branches), purely supersonic outflow ($v > v_*$ branches), or purely subsonic accretion followed by subsonic outflow ($-v_*< v < v_*$ branches), respectively. Red (with $\mathcal{H}=\mathcal{H}_* -0.02$) and green (with $\mathcal{H}=\mathcal{H}_* -0.05$) curves are not real physical behaviors of the fluid. 

The most interesting solution of the fluid motion is depicted by the black curves in Fig.~\ref{ultra-relativistic}, which exhibits transonic behaviors of the fluid outside the black hole horizon. For $v<0$, there are two black hole curves go through the sonic point $(r_*, -v_*)$. One solution starts at the spatial infinity with sub-sonic flow followed by a supersonic flow after it crosses the sonic point, which corresponds to the standard nonrelativistic accretion considered by Bondi in \cite{bondi1952}. Another solution, which starts at the spatial infinity with supersonic flow but becomes sub-sonic after it crosses the sonic point, is unstable according to the analysis presented in \cite{Ahmed:2016cuy}, so such behaviors is very difficult to achieve. For $v>0$, there are two solutions as well. One solution, which starts at horizon with supersonic flow followed a sub-sonic flow after it crosses the sonic point, corresponds to the transoinc solution of the stellar wind, as discussed in \cite{bondi1952} for the non-relativistic accretion. Another solution, similar to the $v<0$ case, is unstable and too hard to achieve \cite{Ahmed:2016cuy}. 

\red{Here we would like to add several remarks about the physical explanations of the flows in Fig.~\ref{ultra-relativistic} with different values of Hamiltonian $\mathcal{H}$. In general, different values of the Hamiltonian represent different initial states of the dynamical system. For the transonic solution of the ultra-relativistic fluid, its Hamiltonian can be evaluated at the sonic point. The Hamiltonian with values different from the transonic one does not represent any transonic solutions of the flow. For examples, the green curve shows the subcritical fluid flow since such flows do not pass through critical point and fail to reach the critical point. In fact, such solutions have a turning or bouncing point which is the nearest point reachable by such fluids beyond which they are bounced back or turned around to infinity. A similar explanation holds for the red curves. The curves shown in blue and magenta can be termed super-critical flows. Although such fluids also do not go through critical point, they already possess velocities more than the allowed critical value. Such flows end up entering the black horizon. It is also worth mentioning that the similar analysis also applies to other fluids including the radiation, sub-relativistic fluid, and polytropic fluid.}

\begin{table}[tbp]
	\centering
	\begin{tabular}{|c|c|c|c|c|c|c|}
		\hline
		\textrm{$\epsilon$} & $0$ &$0.1$ & $0.2$ & $0.3$ & $0.4$ & $0.5$ \\
		\hline
		$r_0$ &$2$ & $1.81818$ & $1.66667$ & $1.53846$ &$1.42857$ & $1.33333$ \\
		\hline
		$r_*$& $2.49998$ & $2.30139$ & $2.14917$ & $2.04111$ & $1.97876$& $1.96016$ \\
		\hline
		$v_* $& $0.70711$ &$ 0.70711$ & $ 0.70711$ & $ 0.70711$ & $0.70711$& $0.70711$ \\
		\hline
		$\mathcal{H}_*$& $0.14311$ &$0.16034$ & $0.17465$ & $0.18507$ & $0.19098$ & $0.19271$ \\
		\hline
	\end{tabular}
	\caption{\label{tab3}
		Values of $r_*$, $v_*$, and $\mathcal{H}_*$ at the sonic point with different values of the black hole parameter $\epsilon$ for the ultra-relativistic fluid $w=1/2$. We use $M=1$ and $a_0 = 0.0001$ in the calculation.}
\end{table}

\subsubsection{Solution for radiation fluid ($w=1/3$)}

\begin{figure}
	\centering
	\includegraphics[width=3.4in]{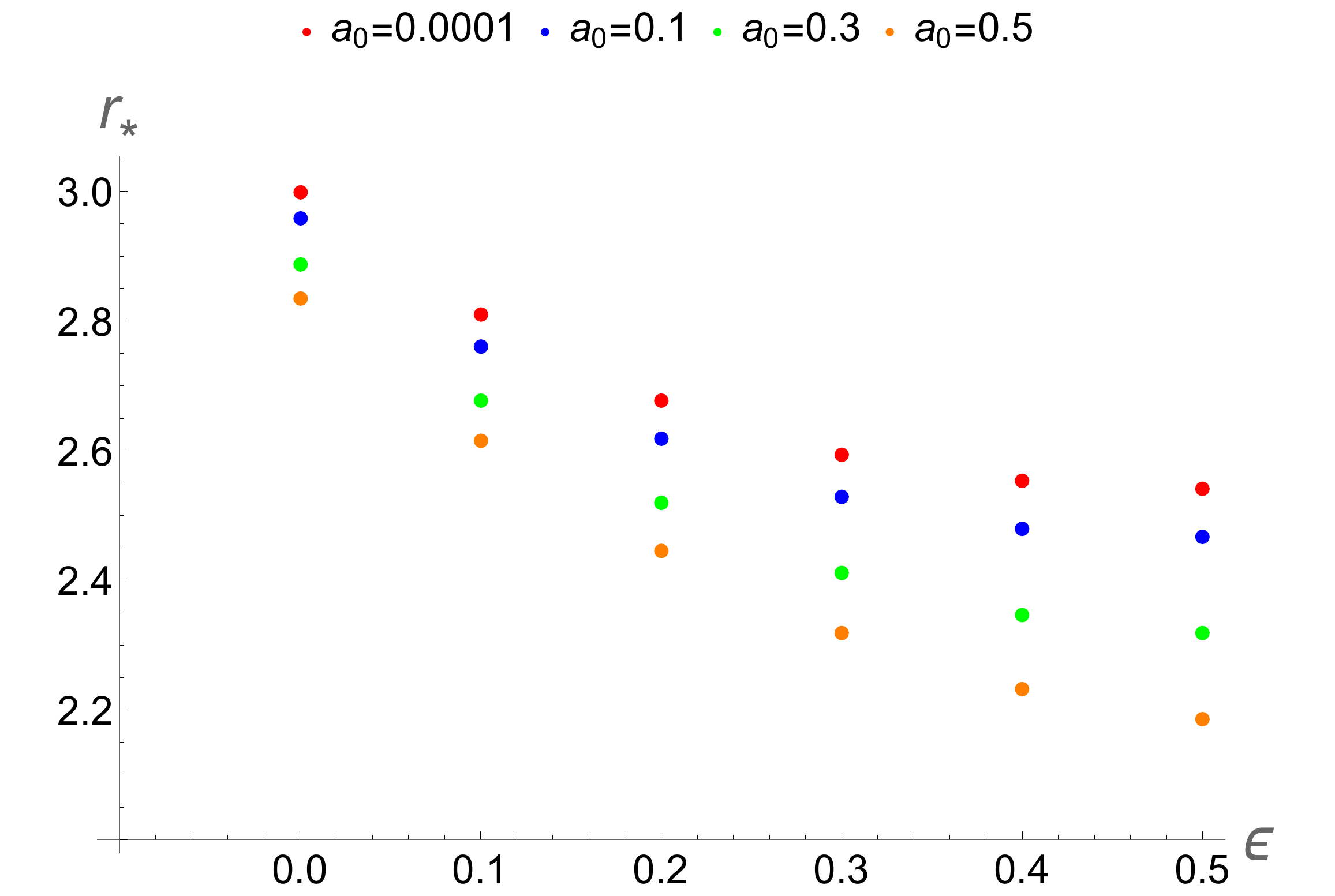}
	\caption{Relation of $r_*$ and $\epsilon$ for different $a_0$ in the spherical accretion process for radiation fluid($w=1/3$).}
	\label{rc(w=1/3)}
\end{figure}

For the radiation fluid, the equation of state is $w=1/3 $. In this case, the Hamiltonian (\ref{H(w)}) becomes
\bqn\lb{H(w=1/3)}
\mathcal{H}
= \frac{\left(1-\frac{2M}{r(1+\epsilon)}\right)^{2/3} \left[1+\frac{4 M^2 (a_0- \epsilon)}{r^2 (1+\epsilon)^2} - \frac{2 M \epsilon }{r(1+\epsilon)}\right]^{2/3}}{r^{4/3}|v|^{2/3} (1-v^2)^{2/3} }, 
\eqn
and then the two-dimensional dynamical system (\ref{DY1}, \ref{DY2}) is
\bqn
\dot{r} =  &&\frac{4 v^{1/3} \left(1 - \frac{2 M}{r (1 +\epsilon)}\right)^{2/3} \left[1 + \frac{
		4 M^2 (a_0 - \epsilon)}{r^2 (1 + \epsilon)^2} - \frac{
		2 M \epsilon}{r (1 + \epsilon)}\right]^{2/3}}{3 r^{
		4/3} (1 - v^2)^{5/3}}\nb\\ &&-\frac{2 \left(1 - \frac{2 M}{r (1 +\epsilon)}\right)^{2/3} \left[1 + \frac{
		4 M^2 (a_0 - \epsilon)}{r^2 (1 + \epsilon)^2} - \frac{
		2 M \epsilon}{r (1 + \epsilon)}\right]^{2/3}}{3 r^{
		4/3}|v|^{5/3} (1 - v^2)^{5/3}}, \lb{r13}
\eqn
\bqn
\dot{v}= &&\frac{4\left(1-\frac{2M}{r(1+\epsilon)}\right)^{2/3} \left[1+\frac{4 M^2 (a_0- \epsilon)}{r^2 (1+\epsilon)^2} - \frac{2 M \epsilon }{r(1+\epsilon)}\right]^{2/3}}{3 r^{7/3}|v|^{2/3}(1-v^2)^{2/3} } \nb\\
&& - \frac{2\left(1-\frac{2M}{r(1+\epsilon)}\right)^{} \left[-\frac{8 M^2 (a_0- \epsilon)}{r^3 (1+\epsilon)^2} + \frac{2 M \epsilon }{r^2 (1+\epsilon)}\right]^{}}{ LT_3} \nb\\ &&-\frac{4 M (1 + \frac{4 M^2 (a_0 - \epsilon)}{r^2 (1 + \epsilon)^2} - \frac{2 M\epsilon}{r (1 + \epsilon)})}{r^2 (1 + \epsilon) LT_3}, \lb{v13},
\eqn
where
\bqn
LT_3=&&3 r^{4/3} |v|^{2/3} (1-v^2)^{2/3} \left(1-\frac{2M}{r(1+\epsilon)}\right)^{1/3}\left[1+\frac{4 M^2 (a_0- \epsilon)}{r^2 (1+\epsilon)^2} - \frac{2 M \epsilon }{r(1+\epsilon)}\right]^{1/3}.\nb\\
\eqn
The sonic points can be found by solving the above two-dimensional dynamical system, when both right hand sides of Eq.~(\ref{r13}) and Eq.~(\ref{v13}) vanish. The values of critical radius $r_*$, sound speed $\pm v_*$, and critical $\mathcal{H}_*$ for different values of $\epsilon$ is summarized in Table.~\ref{tab4} with $w=1/3$, $M=1$, and $a_0 = 0.0001$. Similar to the ultra-relativistic fluid, the critical radius decreases as the increasing of the values of $\epsilon$, while the critical Hamiltonian $\mathcal{H}_*$ increases. We also illustrate the behaviors of the critical radius $r_*$ for radiation fluid with respect to $\epsilon$ for different values of $a_0$ in Fig.~\ref{rc(w=1/3)}. 

\begin{figure}
	\centering
	\includegraphics[width=3.4in]{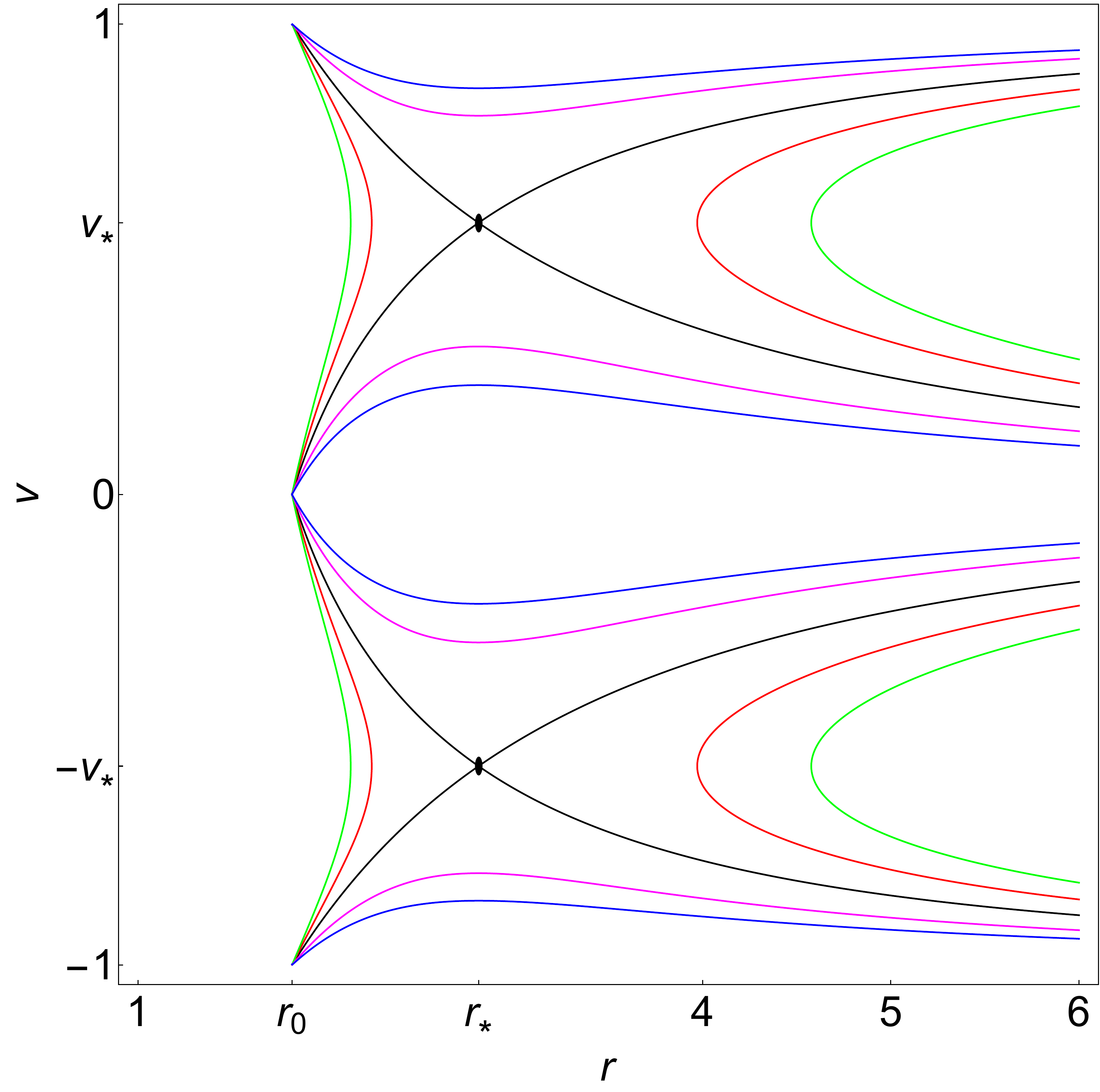}
	\caption{Phase space portraits of the dynamical system (\ref{DY1}, \ref{DY2}) for the radiation fluid ($w=1/3$) with black hole  $M=1$, $\epsilon=0.1$, and $a_0=0.0001$. The critical (sonic) points $(r_*, \pm v_*)$ of this dynamical system are presented by the black spots in the figure. The five curves with color in black, red, green, magenta, and blue correspond to values of Hamiltonian $\mathcal{H}=\mathcal{H}_*,\; \mathcal{H_*}-0.03, \mathcal{H}_*- 0.05, \mathcal{H}+0.05, \mathcal{H}_* + 0.1$, respectively.
	}
	\label{radiation}
\end{figure}

The phase space portraits of this dynamical system for the radiation fluid with $M=1$, $a_0=0.001$, and $\epsilon=0.1$ is displayed in Fig.~\ref{radiation}, in which the physical flows of the radiation fluid in the general parameterized black hole are represents by several curves. One can see that both the critical points in Fig. \ref{radiation}, $(r_*, v_*)$ and $(r_*, -v_*)$, are saddle points of the dynamical system. From Fig.~\ref{radiation}, one also observes that the radiation fluid shares the same types of fluid motion ($w=1/3$) as that for the ultra-relativistic fluid ($w=1/2$) as presented in Fig.~\ref{ultra-relativistic}. Similar to Fig.~\ref{ultra-relativistic}, the magenta and blue curves represent the supersonic flows for $v<-v_*$ or $v>v_*$, while they correspond to sub-sonic flows if $-v_* < v< v_*$. The transonic solutions are presented by the black curves. For $v<0$, one of black curves, which starts at spastically infinity with sub-sonic flow and then becomes supersonic after it crosses the sonic point $(r_*, -v_*)$, corresponds to the standard transonic accretion, and another black curve represent an unstable solution. For $v>0$, one black curve corresponds to the transonic outflow of wind and another one represent an unstable flow, similar to the cases for ultra-relativistic fluid. The green and red curves are unphysical solutions. 

\begin{table}[tbp]
	\centering
	\begin{tabular}{|c|c|c|c|c|c|c|}
		\hline
		\textrm{$\epsilon$} & $0$ &$0.1$ & $0.2$ & $0.3$ & $0.4$ & $0.5$ \\
		\hline
		$r_0$ &$2$ & $1.81818$ & $1.66667$ & $1.53846$ &$1.42857$ & $1.33333$ \\
		\hline
		$r_*$ &$2.99996$ & $2.8096$ & $2.67692$ & $2.59413$ &$2.55246$ & $2.54108$ \\
		\hline
		$ v_* $& $0.57735$ &$0.57735$ & $0.57735$ & $0.57735$ & $0.57735$& $ 0.57735$ \\
		\hline
		$\mathcal{H}_*$& $0.20999$ &$0.22082$ & $0.22845$ & $0.23316$ & $0.2355$ & $0.23613$ \\
		\hline
	\end{tabular}
    \caption{\label{tab4} Values of $r_*$, $v_*$, and $\mathcal{H}_*$ at the sonic point with different values of black hole parameter  $\epsilon$ for the radiation fluid with $w=1/3$. We use $M=1$ and $a_0 = 0.0001$ in the calculation.}
\end{table}

\subsubsection{Solution for sub-relativistic fluid ($w=1/4$)}

\begin{figure}
	\centering
	\includegraphics[width=3.4in]{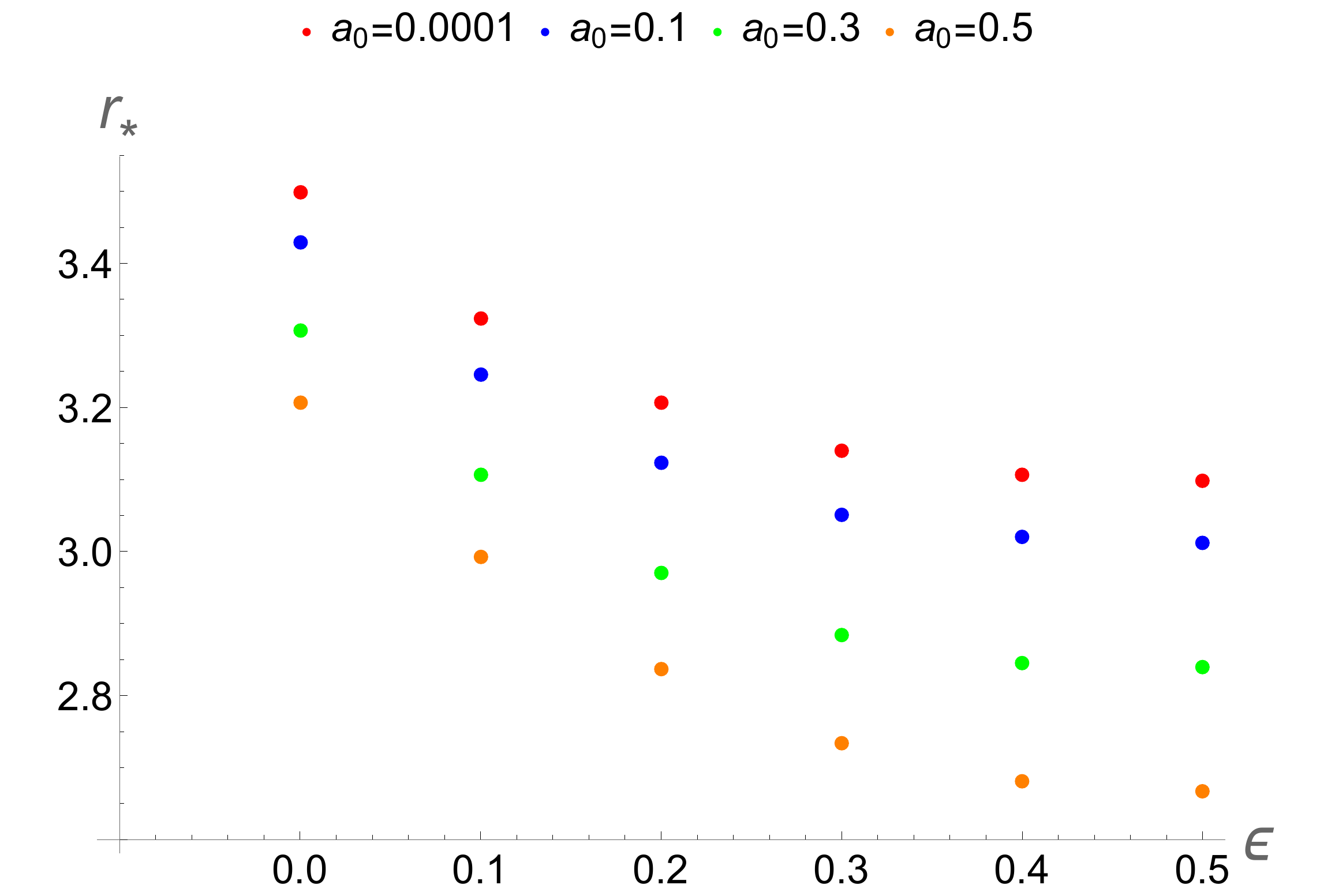}
	\caption{Relation of $r_*$ and $\epsilon$ for different $a_0$ in the spherical accretion process for sub-relativistic fluid($w=1/4$).}
	\label{rc(w=1/4)}
\end{figure}

Let us turn to consider the sub-relativistic fluid, \red{whose energy
	density exceeds their isotropic pressure}, with the equation of state $w = \frac{1}{4}$. In this case, the Hamiltonian (\ref{H(w)}) takes the form
\bqn\lb{H(w=1/4)}
\mathcal{H} =\frac{\left(1-\frac{2M}{r(1+\epsilon)}\right)^{3/4} \left[1+\frac{4 M^2 (a_0- \epsilon)}{r^2 (1+\epsilon)^2} - \frac{2 M \epsilon }{r(1+\epsilon)}\right]^{3/4}}{r\sqrt{|v|}(1-v^2)^{3/4}} ,
\eqn
and then the two-dimensional dynamical system is
\bqn
\dot{r} = &&\frac{3\sqrt{ |v|} \left(1 - \frac{2 M}{r (1 +\epsilon)}\right)^{3/4} \left[1 + \frac{
		4 M^2 (a_0 - \epsilon)}{r^2 (1 + \epsilon)^2} - \frac{
		2 M \epsilon}{r (1 + \epsilon)}\right]^{3/4}}{2r^{} (1 - v^2)^{7/4}}\nb\\ &&-\frac{ \left(1 - \frac{2 M}{r (1 +\epsilon)}\right)^{3/4} \left[1 + \frac{
		4 M^2 (a_0 - \epsilon)}{r^2 (1 + \epsilon)^2} - \frac{
		2 M \epsilon}{r (1 + \epsilon)}\right]^{3/4}}{ r^{}|v|^{3/2} (1 - v^2)^{3/4}},
\eqn
\bqn
\dot{v} = &&\frac{\left(1-\frac{2M}{r(1+\epsilon)}\right)^{3/4} \left[1+\frac{4 M^2 (a_0- \epsilon)}{r^2 (1+\epsilon)^2} - \frac{2 M \epsilon }{r(1+\epsilon)}\right]^{3/4}}{ r^{2}\sqrt{|v|}(1-v^2)^{3/4} } \nb\\
&& - \frac{3\left(1-\frac{2M}{r(1+\epsilon)}\right)^{} \left[-\frac{8 M^2 (a_0- \epsilon)}{r^3 (1+\epsilon)^2} + \frac{2 M \epsilon }{r^2 (1+\epsilon)}\right]^{}}{ LT_4} \nb\\ &&-\frac{4 M (1 + \frac{6 M^2 (a_0 - \epsilon)}{r^2 (1 + \epsilon)^2} - \frac{2 M\epsilon}{r (1 + \epsilon)})}{r^2 (1 + \epsilon) LT_4},
\eqn
where
\bqn
LT_4=&&4 r \sqrt{|v|}  (1-v^2)^{3/4} \left(1-\frac{2M}{r(1+\epsilon)}\right)^{1/4}\left[1+\frac{4 M^2 (a_0- \epsilon)}{r^2 (1+\epsilon)^2} - \frac{2 M \epsilon }{r(1+\epsilon)}\right]^{1/4}.
\eqn
For this dynamical system, similar to the above two cases, we present the values of $r_*$, $v_*$, and $\mathcal{H}_*$ with different values of $\epsilon$ in Table.~\ref{tab5} for $w=1/4$, $M=1$, and $a_0 = 0.0001$. We also plot the behaviors of the critical radius $r_*$ for the sub-relativistic fluid with respect to the black hole parameter $\epsilon$ with different values of $a_0$, \red{as shown in Fig.~{\ref{rc(w=1/4)}}}. It is shown that the critical radius $r_*$ decreases with the increasing of the parameters $\epsilon$ and $a_0$. 

\begin{figure}
	\centering
	\includegraphics[width=3.4in]{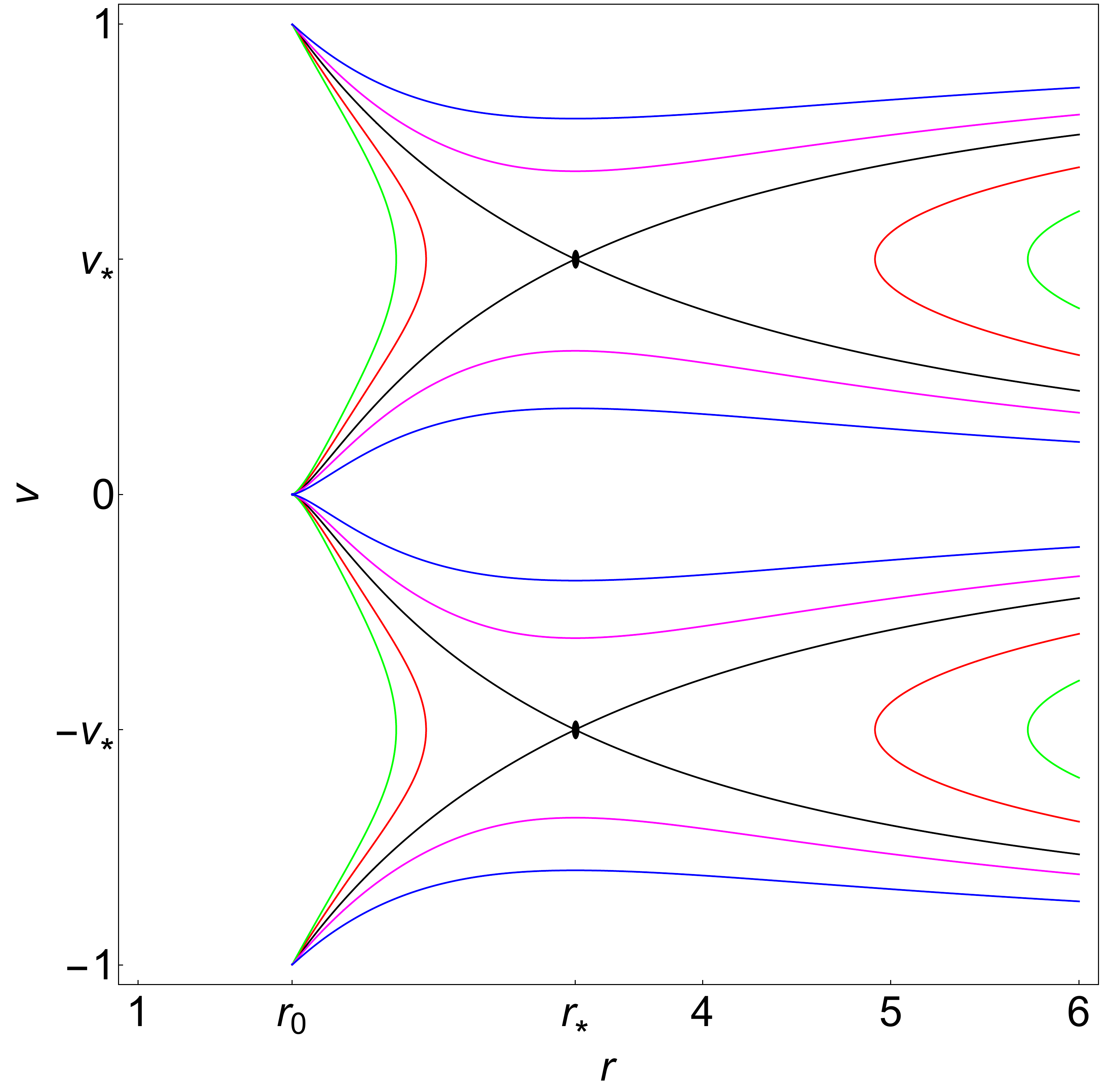}
	\caption{Phase space portraits of the dynamical system (\ref{DY1}, \ref{DY2}) for the sub-relativistic fluid ($w=1/4$) with black hole  $M=1$, $\epsilon=0.1$, and $a_0=0.0001$. The parameters are
		$r_0 \simeq 1.81818$,  $r_* \simeq 3.32303$, $v_* \simeq 0.5$. Black plot: the
		solution curve through the saddle CPs ($r_*, v_*$) and ($r_*, -v_*$) for which
		$\mathcal{H} = \mathcal{H}_* \simeq 0.272806$. Red plot: the solution curve for which $\mathcal{H} = \mathcal{H}_*-  0.03$. Green plot: the solution curve for which $\mathcal{H} = \mathcal{H}_*-  0.05$, Magenta plot: the solution curve for which $\mathcal{H} = \mathcal{H}_* + 0.03$. Blue plot:
		the solution curve for which $\mathcal{H} = \mathcal{H}_* + 0.1$. }
	\label{sub}
\end{figure}

The phase space portraits of the dynamical system for the sub-relativistic fluid is depicted in Fig.~{\ref{sub}}. From this figure, we observe that there are the same types of fluid motion for sub-relativistic fluid ($w=  1/4$) as that for the ultra-relativistic fluid ($w= 1/2$) and the radiation fluid ($w=1/3$). For $v>v_*$, the magenta and blue curves are purely supersonic outflows, while for $v<-v_*$, they represent supersonic accretions. When $-v_* < v < v_*$, there curves are subsonic flows. The black curves shown in Fig.~\ref{sub} are more interesting since they represent the transonic solution of the spherical accretion for $v<0$ and spherical outflow for $v>0$ around the black hole. Similar to the ultra-relativistic fluid and radiation fluid, the red and green curves are unphysical solutions. 

\begin{table}[tbp]
	\centering
	\begin{tabular}{|c|c|c|c|c|c|c|}
		\hline
		\textrm{$\epsilon$} & $0$ &$0.1$ & $0.2$ & $0.3$ & $0.4$ & $0.5$ \\
		\hline
		$r_0$ &$2$ & $1.81818$ & $1.66667$ & $1.53846$ &$1.42857$ & $1.33333$ \\
		\hline
		$r_*$ &$3.49993$ & $3.32303$ & $3.20741$ & $3.13980$ &$3.10743$ & $3.09881$ \\
		\hline
		$v_* $& $0.5$ &$0.5$ & $0.5$ & $0.5$ & $ 0.5$& $ 0.5$ \\
		\hline
		$\mathcal{H}_*$& $0.26557$ &$0.27281$ & $0.27750$ & $0.28022$ & $0.2815$ & $0.28184$ \\
		\hline
	\end{tabular}
	\caption{\label{tab5} Values of $r_*$, $v_*$, and $\mathcal{H}_*$ at the sonic point with different values of the black hole parameter $\epsilon$ for the sub-relativistic fluid $w=1/4$. The black hole parameters $M$ and $a_0$ are set to $M=1$ and $a_0 = 0.0001$.}
\end{table}

\subsection{Polytropic test fluid}

The state of polytropic test fluid can be described by
\bqn
p = \kappa n^\gamma ,
\eqn
where $\kappa$ and $\gamma$ are constants. For ordinary matter, one generally works with the constraint $\gamma > 1$. Following \cite{Ahmed:2015tyi}, we obtain the following expressions of the specific enthalpy
\bqn
h = m +\frac{\kappa \gamma n^{\gamma -1}}{\gamma - 1},
\eqn
where a constant of integration has been identified with the baryonic mass $m$. The three-dimensional speed of sound is given by
\bqn \lb{Y}
c_s^2 = \frac{(\gamma -1) Y}{m(\gamma -1) + Y}~~~(Y \equiv \kappa \gamma n^{\gamma -1}).
\eqn
Using Eq.~(\ref{n^2}) in Eq.~(\ref{Y}), we obtain
\bqn\lb{h}
h = m\left[ 1+ Z \left(\frac{1-v^2}{r^4 N^2 v^2}\right)^{(\gamma -1)/2}\right],
\eqn
where
\bqn
Z \equiv \frac{\kappa \gamma}{m(\gamma -1)} \left| C_1 \right|^{\gamma - 1} = \text{const}.>0,
\eqn
and $Z$ is a positive constant. If the critical points exist, $Z$ takes the special form
\bqn\lb{Z(n_*)}
Z \equiv \frac{\kappa \gamma n^{\gamma -1}_*}{m(\gamma -1)} \left(\frac{r^5_* N^2_{*,r_*}}{4}\right)^{(\gamma - 1)/2}  = \text{const}.>0 .
\eqn
The constant $Z$ depends on the black hole parameters and the test fluid. From Eq. (\ref{Z(n_*)}), it is clear to see that $Z$ is roughly proportional to $\kappa n_* / m $ for a given black hole solution and certain test fluids. 

Inserting Eq.~(\ref{h}) into Eq.~(\ref{H(r,v)}), we evaluate the Hamiltonian by
\bqn\lb{H(N)}
\mathcal{H} = \frac{N^2}{1 - v^2}\left[1 + Z \left( \frac{1 - v^2}{r^4 N^2 v^2}\right)^{(\gamma -1)/2}\right]^2,
\eqn
where $m^2$ has been absorbed into a redefinition of $(\bar{t},\mathcal{H})$. Obviously, $N^2(r) >0 $ and $ N^2_{,r} > 0$ for all $r$. This means that the constant $Z >0 $ (recall that $\gamma >1 $). It is easy to find that there are no global solutions, since the Hamiltonian has remain constant on a solution curve.

Notice that since $ \gamma > 1$, the solution curves do not cross the $r$ axis at points where $v = 0$ and $ r \ne r_0 $, otherwise the Hamiltonian (\ref{H(N)}) would diverge there. The point on the $r$ axis which curves may cross is $(r_0,0)$ only. The horizon $r=r_0$ is a single root to $ N^2(r)=0 $, in the vicinity of which $v$ behaves as
\bqn
\left|v\right| \propto \left| r-r_0\right|^{\frac{2-\gamma}{2(\gamma -1)}}.
\eqn
We see that only solutions with $1<\gamma< 2$ may cross the $r$ axis. Here $\mathcal{H}(r_0,0)$ is the limit of $\mathcal{H}(r,v)$ as $(r,v)$ $\to$ $(r_0,0)$. When $1<\gamma< 2$, the pressure $p = \kappa n^\gamma$ diverges at the horizon as
\bqn
p \propto \left| r-r_0\right|^{\frac{-\gamma}{2(\gamma -1)}}.
\eqn 

Then, inserting 
\bqn
Y = m(\gamma-1) Z \left(\frac{1-v^2}{r^4 N^2 v^2}\right)^{(\gamma -1)/2} 
\eqn
into Eq.~(\ref{Y}), we obtain
\bqn\lb{CsZ}
c_s^2 = Z (\gamma -1- c_s^2)   \left(\frac{1-v^2}{r^4 N^2 v^2}\right)^{(\gamma -1)/2} .
\eqn
which along with Eq. (\ref{c_s^2}) takes the form of the following expressions at the critical points \red{($c_s^2(r_*) = v^2(r_*) = v^2_*$)}:
\bqn
\red{c_s^2(r_*)} =&& Z (\gamma -1- v^2_*)   \left(\frac{1-v^2_*}{r^4_* N^2_* v^2_*}\right)^{(\gamma -1)/2},\lb{vc^2_1}
~\\
v^2_*=&&\frac{M [-12 M^2 \epsilon + r_*^2 (1 + \epsilon)^3 +  4 a_0 M (3 M - r_* (1 + \epsilon))]}{LT_1}.
\lb{vc^2_2}
\eqn
\red{where we have used Eq. (\ref{c_s^2}) to get the right-hand side of Eq. (\ref{vc^2_2}).}
 If there are critical points, the solution of this system of equations in $(r_*,v_*)$ provides all the critical points, with a given value of the positive constant $Z$. And one can use the values of critical points to reduce $n_*$ from Eq. (\ref{Z(n_*)}).

\begin{figure}
	\centering
	\includegraphics[width=3.4in]{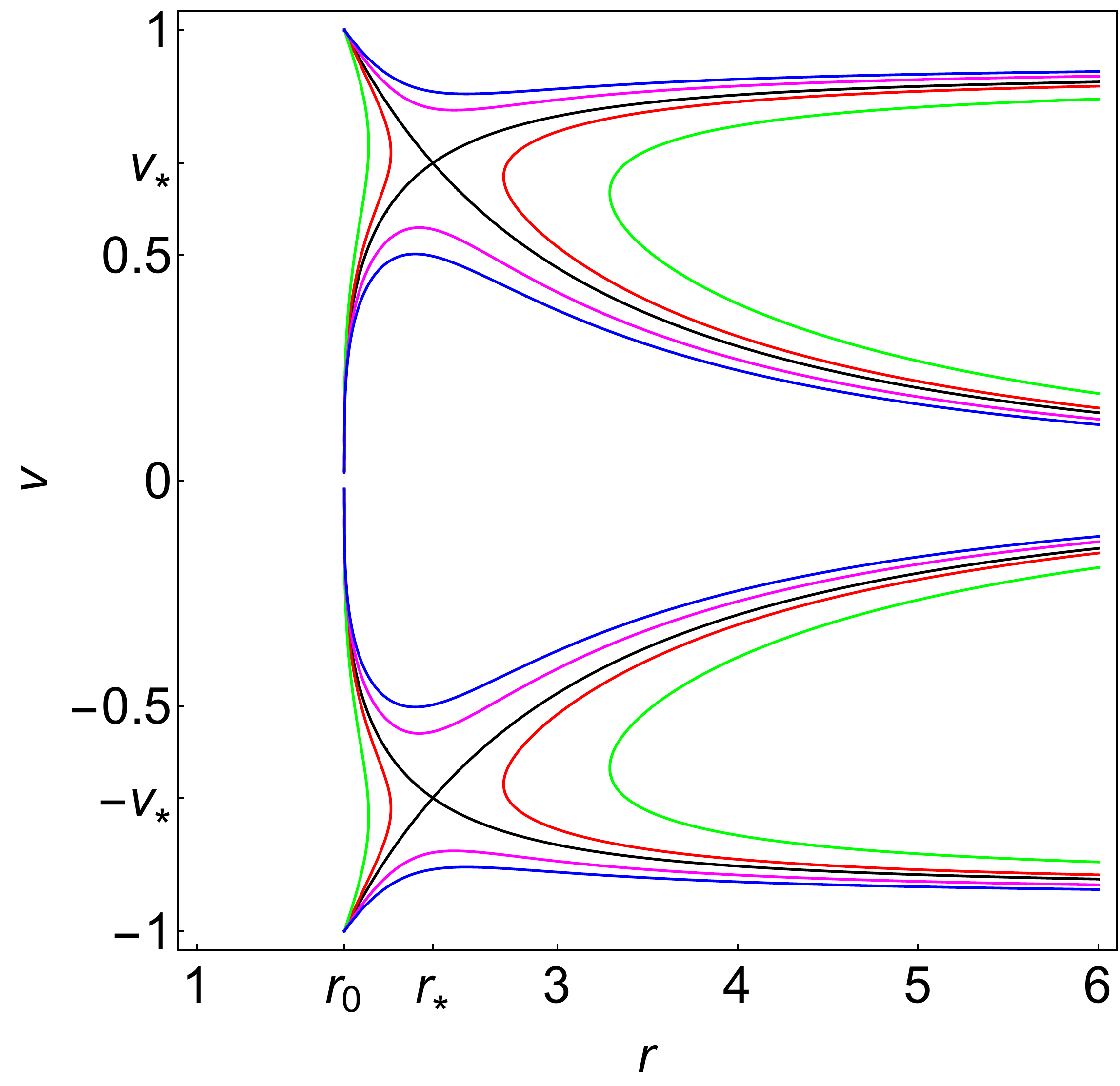}
	\caption{Accretion of a polytropic test fluid. Contour plots of the Hamiltonian (\ref{H(N)}) for $\gamma =5/3$, and $Z=5$, with black hole  $M=1$, $\epsilon=0.1$, and $a_0=0.0001$. The parameters are
		$r_0 \simeq 1.81818$,  $r_* \simeq 2.30998$, $v_* \simeq 0.704098$. Black plot: the
		solution curve through the saddle CPs ($r_*, v_*$) and ($r_*, -v_*$) for which
		$\mathcal{H} = \mathcal{H}_* \simeq 5.5208$. Red plot: the solution curve for which $\mathcal{H} = \mathcal{H}_*-0.3$. Green plot: the solution curve for which $\mathcal{H} = \mathcal{H}_*-1.0$. Magenta plot: the solution curve for which $\mathcal{H} = \mathcal{H}_*+0.5$. Blue plot:
		the solution curve for which $\mathcal{H} = \mathcal{H}_*+1.0$.}
	\label{Polytropic-1}
\end{figure}

Numerical solutions to the dynamical system of Eqs. (\ref{vc^2_1}) and (\ref{vc^2_2}) are shown in Fig. \ref{Polytropic-1}. One can see that there is only one critical point, a saddle point, in accretion ($-1<v<0$) of a polytropic test fluid. And the types of motion of polytropic test fluids, as shown in Fig. \ref{Polytropic-1}, are the same as the types of motion of isothermal test fluids with $w= 1/2$ (c.f. Fig. \ref{ultra-relativistic}), $w=1/3$ (c.f. Fig. \ref{radiation}), and $w=1/4$ (c.f. Fig.\ref{sub}). 

\section{Correspondence Between Sonic Points of Photon Gas and Photon Sphere}
\renewcommand{\theequation}{6.\arabic{equation}} \setcounter{equation}{0}

Recently, it was shown that there is a correspondence between the sonic points of ideal photon gas and the photon sphere in static spherically symmetric spacetimes \cite{Koga:2016jjq}. This important result is valid not only for spherical accretion of the ideal photon gas, but also for rotating accretion in static spherically symmetric spacetimes \cite{Cvetic,Koga:2019teu, Koga:2018ybs}. In this section, we establish this correspondence in the parameterized spherically symmetric black hole.

Let us first consider the the spherical accretion of the ideal photon gas and derive the corresponding sonic points. The equation of state of ideal photon gas in $d$ dimensional space is
\bqn
h = \frac{k \gamma}{\gamma -1} n^{\gamma -1},
\eqn
with
\bqn
\gamma = \frac{d+1}{d},
\eqn
where $k$ is a constant of the entropy\cite{Koga:2016jjq}. And the sound speed of ideal photon gas is constant
\bqn
c_s^2 \equiv \frac{d \ln h }{d \ln n} = \gamma -1.
\eqn
In the general parameterized spherically symmetric spacetimes ($d = 3$), the EoS of ideal photon gas becomes
\bqn
h = 4 k n^{1/3},
\eqn
and the sound speed of ideal photon gas is $c_s^2 = 2$. For the accretion of ideal photon gas in parameterized spherically symmetric spacetimes, the radius $r_*$ of a sonic point is specified
by
\bqn\lb{SPoPG}
\frac{d }{d r}\left(\frac{N}{r}\right) = 0.
\eqn

To proceed let us derive the photon sphere by analyzing the evolution of the photon in the parameterized spherically symmetric black hole. The photon follows the null geodesics in a given black hole spacetime. As the spacetime is spherically symmetric, we can do the calculations in the equatorial plane $\theta=\pi/2$. In order to find the null geodesics around the black hole we can use the Hamilton-Jacobi equation given as follows
\bqn
\frac{\partial S}{\partial \lambda}=-\frac{1}{2}g^{\mu\nu}\frac{\partial S}{\partial x^\mu}\frac{\partial S}{\partial x^\nu},
\eqn
where $\lambda$ is the affine parameter of the null geodesic and $S$ denotes the Jacobi action of the photon. The Jacobi action $S$ can be separated in the following form,
\begin{equation}
S=-Et+L\phi+S_r(r),
\end{equation} 
where $E$ and $L$ represent the energy and angular momentum of the photon respectively. The function $S_r(r)$ depends only on $r$. 

Now substituting the Jacobi action into the Hamilton-Jacobi equation, we obtain
\bqn
S_r(r)&=&\int^r\frac{B^2(r)\sqrt{R(r)}}{r^2 N^2(r)}dr,
\eqn
where
\begin{eqnarray}
R(r) &=&-\frac{r^2 N^2(r)L^2}{B^2(r)}+\frac{r^4 E^2}{B^2(r)}.
\end{eqnarray}
Then variation of the Jacobi action gives the following equations of motion for the evolution of the photon,
\bqn
\frac{dt}{d\lambda} &=&\frac{E}{N^2(r)},\\
\frac{d\phi}{d\lambda} &=&\frac{L}{r^2},\\
\frac{dr}{d\lambda} &=&\frac{\sqrt{R(r)}}{r^2}.
\eqn

To determine the radius of the photon sphere of the black hole, we need to find the critical circular orbit for the photon, which can be derived from the unstable condition
\bqn
R(r)=0,\qquad \frac{dR(r)}{dr}=0.\lb{ZZZ}
\eqn
For parameterized spherically symmetric black hole, from above conditions one finds
\bqn\lb{Photo S}
\frac{d}{dr}\left(\frac{N}{r}\right)=0.
\eqn
This is the condition for determining the radius of the photon sphere. One can see that Eq. (\ref{SPoPG}) is actually same as Eq. (\ref{Photo S}), which means that the critical radius $r_*$ of a sonic point, for the accretion of ideal photon gas in parameterized  spherically symmetric spacetimes, is equal to the radius of the photon sphere. 

\begin{figure}
	\centering
	\includegraphics[width=3.4in]{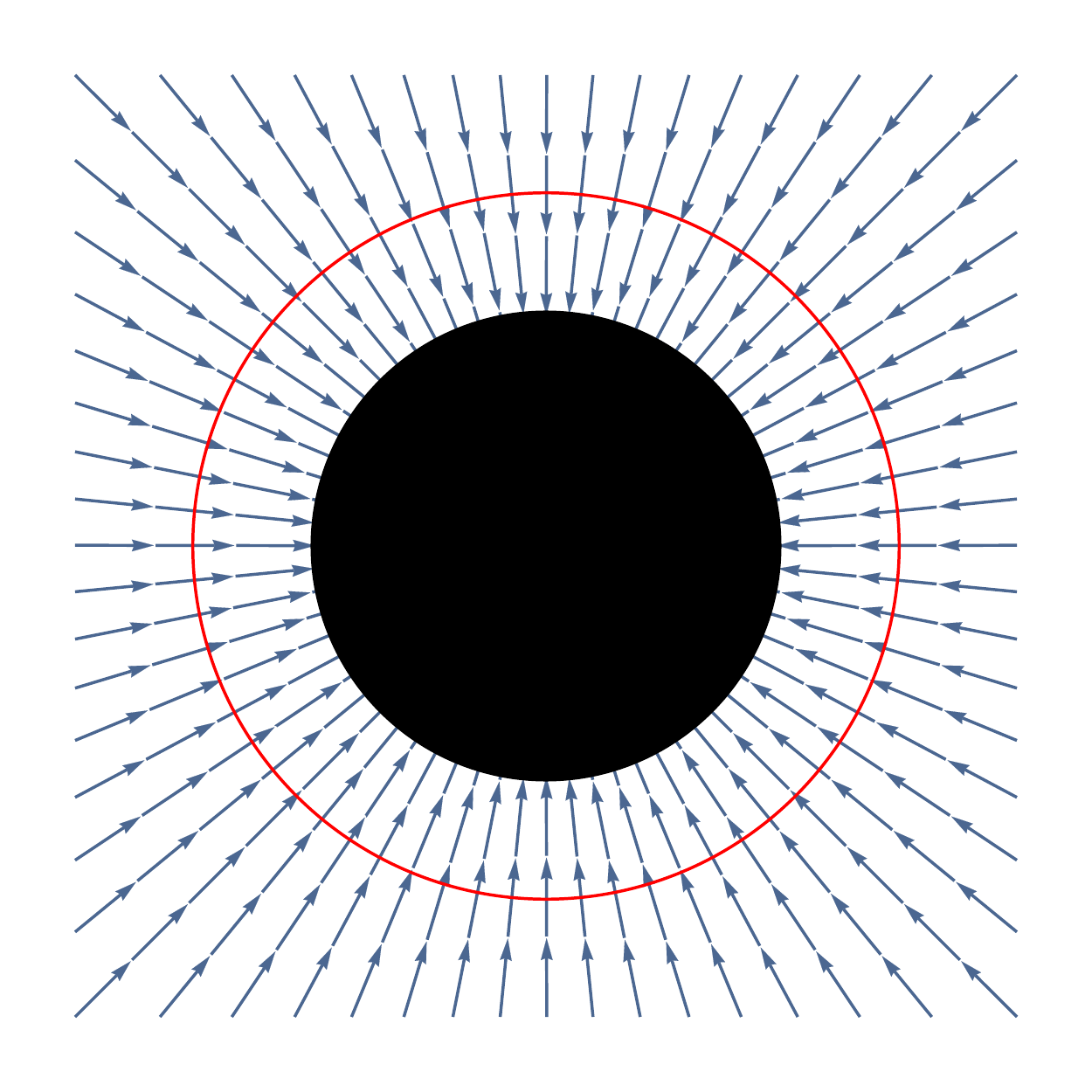}
	\caption{Schematic plot of the spherical accretion of the ideal photon gas onto a spherically symmetric black hole and its photon sphere (the red circle). The red circle has two folds meanings, since it represents both the photon sphere and the sonic radius of the spherical accreting of the ideal photon gas.}
	\label{Photosphere}
\end{figure}

With Eq.~(\ref{Photo S}), one can get the critical radius $r_*$ and the radius of photon sphere in parameterized spherically symmetric spacetimes
\bqn
r_* = \left. \left(\frac{1}{N} \frac{d N}{dr} \right)^{-1} \right|_{r=r_*}.
\eqn
By using Eq.~(\ref{N2}) into above equation , one can obtain the expression of $r_*$,
\bqn
r_* = &&M + \frac{(1 + \text{i} \sqrt{3}) M^2 (1 + \epsilon) [-8 a_0 + 3 (1 + \epsilon)^2]}{2LT_{5}} + \frac{(1 - \text{i} \sqrt{3}) LT_{5}}{6 (1 + \epsilon)^3},
\eqn
where
\bqn
LT_{5} = &&[-27 M^3 (1 + \epsilon)^6 (1 + a_0 (6 - 4 \epsilon) - 7 \epsilon + 3 \epsilon^2 + \epsilon^3) + 6 \sqrt{3} \sqrt{LT_{6}
}]^{1/3},
\eqn
and
\bqn
LT_{6}= &&M^6 (1 + \epsilon)^{12} [128 a_0^3 - 9 a_0^2 (-11 + 68 \epsilon + 4 \epsilon^2) \nb\\
&&- 135 \epsilon (1 - 2 \epsilon + 3 \epsilon^2 + \epsilon^3) + 135 a_0 (1 - 3 \epsilon + 7 \epsilon^2 + \epsilon^3)].
\eqn
It is easy to verify that when the extra parameters ($a_0$ and $\epsilon$) are set to zero, the critical radius reduces to $r_*=3M$.  In Fig.~\ref{Photosphere}, we provide a schematic plot of the  spherical accretion of the ideal photon gas onto a spherically symmetric black hole and its photon sphere (represented by the red circle). The red circle in Fig.~\ref{Photosphere} thus has two folds meanings, since it represents both the photon sphere and the sonic radius of the spherical accreting of the ideal photon gas.

\section{\red{Conclusions and Discussions}}
\renewcommand{\theequation}{7.\arabic{equation}} \setcounter{equation}{0}

In this paper, we study the spherical accretion flow of perfect fluid onto the general parameterized spherically symmetric black hole. For this purpose, we first formulate two basic equations for describing the accretion process and present the general formulas for determining the sonic points or critical points. These two equations are derived from the conservation law of energy and particle number of the fluid. With this two equations, we then analyze the accretion process of various perfect fluids, such as the isothermal fluid of ultra-stiff, ultra-relativistic, and sub-relativistic types and polytropic fluid, respectively. The flow behaviors of these test fluids around the general parameterized spherically symmetric black hole are studied in details and shown graphically in Figs.~\ref{ultra}, \ref{ultra-relativistic}, \ref{radiation}, \ref{sub}. For isothermal fluid, it is interesting to mention that the sonic point does not exist only for the ultra-stiff fluid with $w=1$ and thus the transoinc solutions exist for ultra-relativistic fluid with $w=1/2$, radiation fluid with $w=1/3$, and sub-relativistic fluid $w=1/4$ . \red{And the $\epsilon$ will influence $r_0,~r_*,~\mathcal{H}_*$, but not $v_*$, which means the influence of the position of the event horizon.} For polytropic fluid, it is shown in Fig.~\ref{Polytropic-1} graphically that it poses similar flow behaviors as that of the isothermal fluids with $w=1/2$, $w=1/3$, and $w=1/4$. \red{Here we would like to mention that the results presented in this paper can also be reduced to specific cases in several modified theories of gravity. For example, one can map the results here to the first type Einstein-\AE{}ther black hole in \cite{Zhu:2019ura, ding} by setting 
\bqn
\epsilon=\frac{M-\sqrt{M^2-\ae^2}}{M+\sqrt{M^2-\ae^2}},\\
a_0= \frac{\ae^2}{(M+\sqrt{M^2-\ae^2})^2}, \;\; b_0=0,\\
a_i =0, \;\; b_i=0, \;\;\; (i>0),
\eqn
where
\bqn
\ae^2= - \frac{2 c_{13}-c_{14}}{2(1-c_{13})} M^2
\eqn
with $c_{13}$ and $c_{14}$ being the coupling constants in the Einstein-\AE{}ther theory. It is interesting to mention that flow behaviors for different test fluids in this paper are qualitatively consistent with those studied in \cite{Shahzad:2020uzj} for the spherical accretion in the Einstein-\AE{}ther theory. }

We further consider the spherical accretion of the ideal photon gas and derive the radius of its sonic point. Comparing the radius with that of the photon sphere in the general parameterized spherically symmetric black hole, we study the correspondence between the sonic points of accreting photon gas and the photon sphere for the general parameterized spherically symmetric black hole. 

With the above main results, we would like to mention several directions that can be carried out to extend our analysis. First, the spherical accretion is the simplest accretion scenario, in which the accreting matter falls steadily and radially into the black hole. \red{This is an extreme simple case.} Therefore, it is interesting to explore the accreting behaviors of various matter when the spherical symmetry approximation is relaxed by considering a non-zero relative velocity between the black hole and the accreting matter. This scenario is also known as wind accretion or Bondi–Hoyle–Lyttleton accretion \cite{hoyle1, hoyle2, hoyle3} (see \cite{hoyle4} for a review). \red{We will consider the more complicated accretion disk model, which is more related to the real observations, in our future work. }

Second, it is also interesting to extend our analysis to the rotating black holes. In a rotating background, one may consider rotating fluids accreting onto a rotating black hole. The rotation of the fluids can lead to the formation of a disc-like structure around black hole, and such accretion discs are the most commonly studied engines for explaining astrophysical phenomena such as active galactic nuclei, X-ray binaries, and gamma-ray bursts. However, considering rotation brings complications into the accretion problem, in which case the study heavily relies on numerical calculations. 

At last, when one considers rotating black hole, its shadow does not correspond to a photon sphere, but a photon region. An immediate question now arises that what structure in the rotating accretion of idea photon gas corresponds to the photon region of the rotating black hole. This is still an open issue.


\acknowledgments

C.L., T.Z., and Q.W. are supported by National Natural Science Foundation of China with the Grants No.11675143, the Zhejiang Provincial Natural Science Foundation of China under Grant No. LY20A050002, and the Fundamental Research Funds for the Provincial Universities of Zhejiang in China under Grants No. RF- A2019015.



\end{document}